\PassOptionsToPackage{unicode}{hyperref}
\PassOptionsToPackage{hyphens}{url}
\PassOptionsToPackage{dvipsnames,svgnames,x11names}{xcolor}
\documentclass[
  11pt,
]{article}
\usepackage{amsmath,amssymb}
\usepackage{iftex}
\ifPDFTeX
  \usepackage[T1]{fontenc}
  \usepackage[utf8]{inputenc}
  \usepackage{textcomp} 
\else 
  \usepackage{unicode-math} 
  \defaultfontfeatures{Scale=MatchLowercase}
  \defaultfontfeatures[\rmfamily]{Ligatures=TeX,Scale=1}
\fi
\usepackage{lmodern}
\ifPDFTeX\else
\fi
\IfFileExists{upquote.sty}{\usepackage{upquote}}{}
\IfFileExists{microtype.sty}{
  \usepackage[]{microtype}
  \UseMicrotypeSet[protrusion]{basicmath} 
}{}
\makeatletter
\@ifundefined{KOMAClassName}{
  \IfFileExists{parskip.sty}{%
    \usepackage{parskip}
  }{
    \setlength{\parindent}{0pt}
    \setlength{\parskip}{6pt plus 2pt minus 1pt}}
}{
  \KOMAoptions{parskip=half}}
\makeatother
\usepackage{xcolor}
\usepackage[margin=2cm]{geometry}
\usepackage{longtable,booktabs,array}
\usepackage{calc} 
\usepackage{etoolbox}
\makeatletter
\patchcmd\longtable{\par}{\if@noskipsec\mbox{}\fi\par}{}{}
\makeatother
\IfFileExists{footnotehyper.sty}{\usepackage{footnotehyper}}{\usepackage{footnote}}
\makesavenoteenv{longtable}
\usepackage{graphicx}
\makeatletter
\def\maxwidth{\ifdim\Gin@nat@width>\linewidth\linewidth\else\Gin@nat@width\fi}
\def\maxheight{\ifdim\Gin@nat@height>\textheight\textheight\else\Gin@nat@height\fi}
\makeatother
\setkeys{Gin}{width=\maxwidth,height=\maxheight,keepaspectratio}
\makeatletter
\def\fps@figure{htbp}
\makeatother
\providecommand{\tightlist}{%
  \setlength{\itemsep}{0pt}\setlength{\parskip}{0pt}}
\newlength{\cslhangindent}
\newlength{\csllabelwidth}
\newlength{\cslentryspacingunit} 
\newenvironment{CSLReferences}[2] 
 {
  \setlength{\parindent}{0pt}
  \ifodd #1
  \let\oldpar\par
  \def\par{\hangindent=\cslhangindent\oldpar}
  \fi
  \setlength{\parskip}{#2\cslentryspacingunit}
 }%
 {}
\usepackage{calc}

\usepackage{arydshln}              
\usepackage{float}
\usepackage{newunicodechar}
\newunicodechar{→}{\ensuremath{\rightarrow}}
\newunicodechar{≈}{\ensuremath{\approx}}
\newunicodechar{−}{\ensuremath{-}}
\newunicodechar{×}{\ensuremath{\times}}
\ifLuaTeX
  \usepackage{selnolig}  
\fi
\IfFileExists{bookmark.sty}{\usepackage{bookmark}}{\usepackage{hyperref}}
\IfFileExists{xurl.sty}{\usepackage{xurl}}{} 
\hypersetup{
  pdftitle={Gepard: Real-Time Decoder-Only TTS Native to vLLM},
  pdfauthor={Denis Pavlov Ulanbek Abdurazakov Nursultan Bakashov  nineninesix.ai},
  colorlinks=true,
  linkcolor={blue},
  filecolor={Maroon},
  citecolor={Blue},
  urlcolor={Blue},
  pdfcreator={LaTeX via pandoc}}

\title{GEPARD: A GEnerative, Prosody-aware, Autoregressive text-to-speech model for Realtime Dialogue}
\usepackage{etoolbox}
\makeatletter
\providecommand{\subtitle}[1]{
  \apptocmd{\@title}{\par {\large #1 \par}}{}{}
}
\makeatother
\subtitle{Technical Report}
\author{Denis Pavlov \quad Ulanbek Abdurazakov \quad Nursultan Bakashov
\\[2pt] \href{https://www.nineninesix.ai/}{nineninesix.ai}}
\date{2026}

\begin{document}
\maketitle

\hypertarget{abstract}{%
\subsection{Abstract}\label{abstract}}

We present GEPARD, a multilingual, streaming text-to-speech model for realtime spoken dialogue. GEPARD generates speech autoregressively with an LLM backbone — text and audio embeddings are trained jointly within a single model — and decodes the resulting tokens to a waveform with an FSQ-based NanoCodec, streaming audio chunk-by-chunk as text arrives.

Gepard is an autoregressive (decoder-only) TTS model for interactive
voice agents, designed to ensure ultra-low latency and high throughput
under production serving conditions. The central scientific and
practical goal of this study is to develop a TTS architecture that can
be served by standard LLM engines (specifically, vLLM
(\protect\hyperlink{ref-kwon2023vllm}{Kwon et al. 2023})) without
modifying the source code of their compute kernels. This requirement
defines the overarching design principle: the computational backbone is
a standard full-attention transformer, while all non-trivial auxiliary
mechanisms (dynamic voice cloning, text augmentations, classifier-free
guidance) are moved outside the autoregressive generation cycle (decode
loop) or distilled directly into the model's weights.

On streaming end-to-end inference, a single Gepard stream achieves a
Real-Time Factor (RTF) of \(\approx 0.067\) (about 15 times faster than
real-time). Under concurrent load with 256 simultaneous streams, the
system demonstrates an aggregate speedup (xRT) of \(\approx 204\times\)
on a single server-class GPU. This report details: (1) system-level
solutions for vLLM-native serving; (2) the short register (1--2 words)
failure mode as a fundamental issue in speech decoders, along with
methods for its quantitative evaluation and elimination; and (3)
distillation of two-pass classifier-free guidance over text into
single-pass weights via DPO
(\protect\hyperlink{ref-rafailov2023dpo}{Rafailov et al. 2023}).

\begin{center}\rule{0.5\linewidth}{0.5pt}\end{center}

\hypertarget{introduction-and-problem-formulation}{%
\section{1. Introduction and Problem
Formulation}\label{introduction-and-problem-formulation}}

\hypertarget{motivation-and-central-thesis}{%
\subsection{1.1. Motivation and Central
Thesis}\label{motivation-and-central-thesis}}

Interactive voice agents impose unprecedented limits on the latency to
the first audio frame (Time to First Audio, TTFA) and the economic cost
of scaling infrastructure. While most modern autoregressive TTS models
focus solely on maximizing synthesis quality using complex specialized
decoders, this work is governed by a strict product constraint: the
model must operate on top of a standard vLLM engine
(\protect\hyperlink{ref-kwon2023vllm}{Kwon et al. 2023}).

The vLLM engine implements highly efficient continuous batching and
PagedAttention mechanisms for standard LLM architectures
(\protect\hyperlink{ref-kwon2023vllm}{Kwon et al. 2023}). Introducing
custom operations into the internal autoregressive loop (such as
layer-wise depth-transformers over codec codes, cross-attention
mechanisms to the audio interface in intermediate layers, or two-pass
classifier-free guidance at each decoding step) makes it impossible to
use stock vLLM, breaking continuous batching and drastically reducing
system throughput.

The central thesis of our work is as follows: \emph{high-performance
autoregressive speech synthesis can be realized within a standard
language model architecture without reducing serving engine throughput
by moving all custom components to the prefill phase or pre-distilling
them into the weights.}

\hypertarget{overarching-design-principle}{%
\subsection{1.2. Overarching Design
Principle}\label{overarching-design-principle}}

From the thesis formulated in §1.1, the overarching design principle of
the Gepard architecture follows:

\begin{quote}
The computational backbone remains a standard full-attention
transformer. Any non-standard modal transformations are performed once
during the prefill phase (or offline during training data generation) or
baked into the weights during fine-tuning.
\end{quote}

The practical implementation of this principle includes the following
decisions: - \textbf{Zero-shot Voice Cloning} is moved to the prefix of
the input sequence: the speaker representation is extracted by a frozen
audio compressor once during prefill and does not participate in the
step-by-step autoregressive decode loop. - \textbf{Preventing generation
collapse on ultra-short sequences} is addressed via text augmentation
(text repetitions in prefill) and requires no dynamic logic changes at
inference. - \textbf{Classifier-Free Guidance (CFG)}, which requires two
passes (conditional and unconditional) for each generated frame
(\protect\hyperlink{ref-ho2022cfg}{Ho and Salimans 2022}), is eliminated
from the serving phase: its effect is distilled into the weights of a
single-pass model during DPO training using offline paired generations
(\protect\hyperlink{ref-rafailov2023dpo}{Rafailov et al. 2023}).

\hypertarget{inference-speed-and-scaling}{%
\subsection{1.3. Inference Speed and
Scaling}\label{inference-speed-and-scaling}}

The performance evaluation of Gepard was carried out in two stages: 1.
\textbf{Concept validation stage (sanity check):} An early single-stream
vLLM run on a rented RTX 5090 (Vast.ai), without strict environment
controls, demonstrated an RTF of \(\approx 0.040\) and a TTFA of
\(\approx 0.032\text{ s}\) (approximately 25\(\times\) faster than
real-time). This confirmed the viability of the vLLM-native approach. 2.
\textbf{Production serving stage (realistic serving):} Full end-to-end
testing (backbone + neural codec) under concurrent load using the SSE
streaming protocol on server-class GPUs. The results show: - In
single-thread mode, the RTF is \(\approx 0.067\) with a TTFA (TTFB) of
\(\approx 0.046\text{ s}\). - Aggregate throughput scales linearly up to
xRT \(\approx 204\times\) with 256 concurrent streams. - The optimal
operating range is 64--128 streams per GPU, where each stream maintains
comfortable interactive performance (RTF \textless{} 0.75).

The discrepancy in single-stream RTF (0.040 vs.~0.067) reflects the
measurement setup, not a regression: the sanity stage was an early,
uncontrolled single-stream run on a rented RTX 5090, whereas the
production stage is a rigorous end-to-end measurement on the RTX PRO
6000 under the full streaming-and-concurrency harness. Both agree in
order of magnitude and confirm real-time operation with margin.

\hypertarget{scientific-and-practical-contributions}{%
\subsection{1.4. Scientific and Practical
Contributions}\label{scientific-and-practical-contributions}}

The development of Gepard does not aim to achieve absolute SOTA on
intelligibility metrics (on Seed-TTS-eval
(\protect\hyperlink{ref-bytedance2024seedtts}{Anastassiou et al. 2024}),
the model lies in the middle range). The contributions of this work are
focused on the following aspects: 1. \textbf{System-level vLLM-native
solution:} Proof of the feasibility of integrating TTS into standard LLM
serving frameworks while preserving high memory utilization and
throughput. 2. \textbf{``Short Register'' investigation:} Detailed
description and mathematical analysis of the failure mode of short
phrases (1--2 words), where autoregressive TTS decoders tend to run into
infinite loops (runaway) or skip words. We propose diagnostic tools
(entropy and stop-probability probes) and a compensation method. 3.
\textbf{CFG distillation via DPO:} A schema for transferring the
improvements of two-pass generation with CFG into a single-pass model
with LoRA adapters using length- and quality-normalized reward scaling.

\hypertarget{scope-and-assumptions}{%
\subsection{1.5. Scope and Assumptions}\label{scope-and-assumptions}}

\begin{itemize}
\tightlist
\item
  \textbf{Early development stage (version 0.0.1):} The quantitative
  metrics presented serve as a baseline for optimization and will be
  improved in subsequent versions.
\item
  \textbf{English dominance:} Despite the multilingual nature of the
  pretraining (which included Spanish, Portuguese, Dutch, and Kyrgyz),
  the audio interface of Gepard was optimized primarily on English data.
  Synthesis quality in other languages remains limited and requires
  specialized fine-tuning.
\item
  \textbf{Study of representation leakage in cloning:} The current
  implementation of voice cloning serves as a Proof of Concept (PoC).
  The low speaker similarity observed on unseen voices is analyzed in
  detail in §2.6 and §7 as a representation leakage phenomenon in
  quantized embeddings, rather than just a technological shortcoming.
  This similarity is evaluated using WavLM
  (\protect\hyperlink{ref-chen2021wavlm}{Chen et al. 2021}) speaker
  embeddings.
\end{itemize}

\begin{center}\rule{0.5\linewidth}{0.5pt}\end{center}

\hypertarget{architecture}{%
\section{2. Architecture}\label{architecture}}

\hypertarget{overview}{%
\subsection{2.1. Overview}\label{overview}}

Gepard is an autoregressive transformer that takes text as input and
generates discrete audio codes of a neural codec. The model is
decoder-only: it has no \texttt{lm\_head} and no text generation: it
decodes only audio tokens and predicts the end of speech.

The backbone is based on Qwen3.5 (14 blocks, hidden dimension 1024, 8
attention heads, \(\approx 500\text{M}\) parameters); the full model
with the audio interface and voice cloning contains
\(\approx 555.7\text{M}\) parameters. The input sequence of the backbone
is formatted as:

\[
\bigl[\, \text{prefix}_{(K)} \;\big|\; \text{text}_{(T_\text{text})} \;\big|\; \text{audio}_{(T_\text{audio})} \,\bigr],
\]

where \texttt{prefix} denotes the optional \(K = 8\) speaker tokens for
voice cloning (present only when voice cloning is enabled, §2.6), and
the output heads are applied only to the audio region. The codec is
NVIDIA NanoCodec (\protect\hyperlink{ref-nvidia2025nanocodec}{Casanova
et al. 2025}) (21.5 frames/s, 1.89 kbps configuration).

The subsequent subsections follow the processing pipeline: backbone
(§2.2), codec (§2.3), audio interface from tokens to backbone (§2.4),
output heads and loss function (§2.5), and voice cloning (§2.6).

\begin{figure}
\centering
\includegraphics[width=0.92\textwidth,height=\textheight]{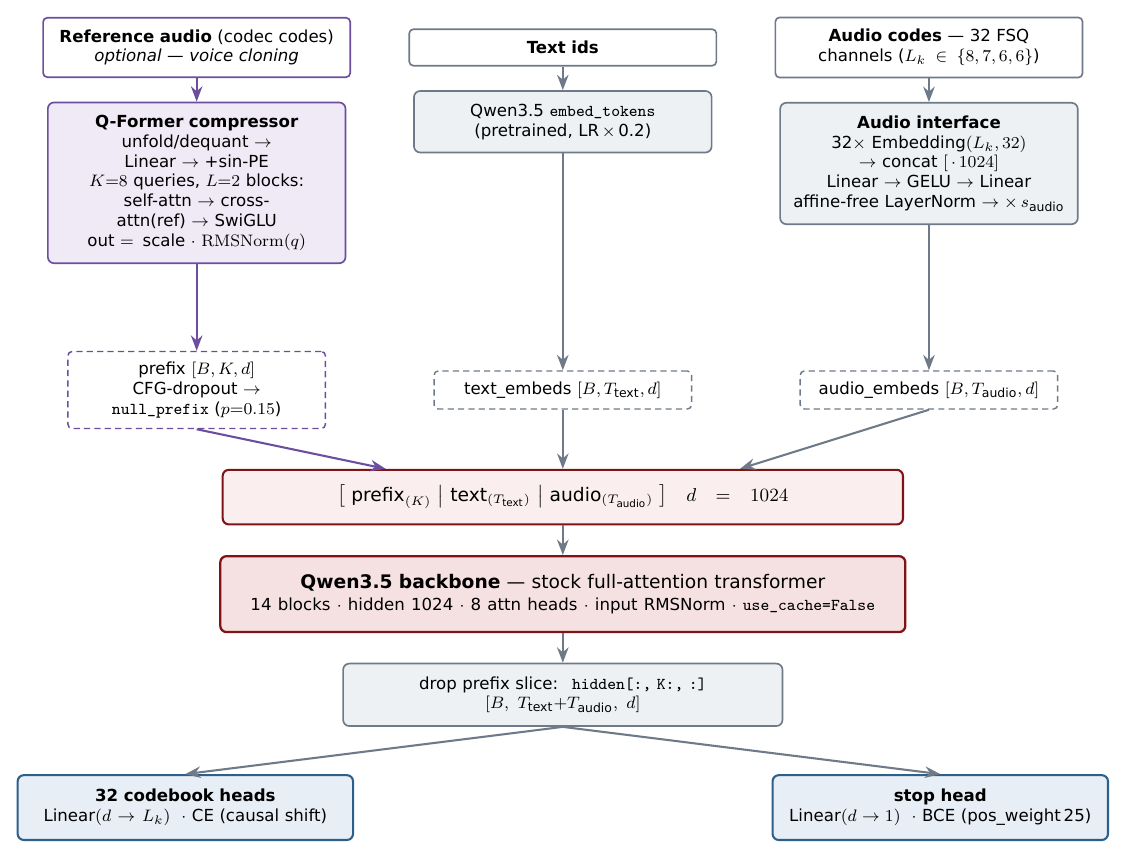}
\caption{Gepard architecture. Three input streams --- the optional
Q-Former voice-cloning prefix, the text embedding, and the audio
interface (32 FSQ channels) --- are concatenated into a single sequence
and processed by a stock full-attention Qwen3.5 backbone. The prefix
slice is dropped before the 32 codebook heads (CE) and the stop head
(BCE).\label{fig:architecture}}
\end{figure}

\hypertarget{backbone}{%
\subsection{2.2. Backbone}\label{backbone}}

The backbone is based on Qwen 3.5. Its ``stock'' nature means that the
backbone remains a standard full-attention transformer without custom
operations, and its structure is not modified during TTS training. The
origin of this base model is clarified by two notes.

First, the \texttt{LinearBlock} layers were removed from the original
Qwen3.5, leaving only classic full attention in all 14 blocks, to ensure
compatibility with standard FlashAttention-2
(\protect\hyperlink{ref-dao2023flashattention2}{Dao 2023}). Only the
text embedding weights and the tokenizer were inherited from the
pretrained Qwen; the rest of the backbone was trained from scratch, as
the removal of \texttt{LinearBlock} altered the layer sequence and made
the old weights incompatible.

Second, the decision to retain only full attention is empirical. It is
based on two informal runs under equal conditions, where the model
without linear layers sounded noticeably better. Quantitative
measurements were not performed, and loss is not indicative here (due to
structural differences), so this decision remains a candidate for
revision. A secondary motivation was that full attention reliably fits
vLLM and simplifies serving.

The backbone runs with \texttt{use\_cache=False} during training; there
are no layer-level hooks or overrides inside it.

\hypertarget{codec-groupfsq-instead-of-residual-vector-quantization-rvq}{%
\subsection{2.3. Codec: GroupFSQ instead of Residual Vector Quantization
(RVQ)}\label{codec-groupfsq-instead-of-residual-vector-quantization-rvq}}

The audio tokenizer is the NVIDIA NeMo NanoCodec, operating at 22 kHz /
21.5 frames/s / 1.89 kbps
(\texttt{nvidia/nemo-nano-codec-22khz-1.89kbps-21.5fps} on Hugging
Face), loaded via NeMo's \texttt{AudioCodecModel}. The sampling rate is
22,050 Hz, the frame rate is 21.5 Hz (about 1024 samples per frame), and
the bitrate is \(\approx 1.89\text{ kbps}\). Our codec utilizes a
GroupFSQ (Finite Scalar Quantization) scheme
(\protect\hyperlink{ref-mentzer2023fsq}{Mentzer et al. 2023}), which
splits the latent space into \(G = 8\) independent groups (subspaces),
each projected into a low-dimensional vector and quantized by its own
FSQ grid with levels \(L = [8, 7, 6, 6]\) (grid capacity is
\(8\cdot7\cdot6\cdot6 = 2016\)).

A key architectural decision in Gepard is the choice of a codec based on
GroupFSQ instead of the Residual Vector Quantization (RVQ) systems
dominant in literature, such as EnCodec
(\protect\hyperlink{ref-defossez2022encodec}{Défossez et al. 2022}),
SoundStream, or DAC (\protect\hyperlink{ref-kumar2023dac}{R. Kumar et
al. 2023}). Comparing these approaches highlights the trade-off between
inference speed and the model's convergence complexity during training:

\begin{enumerate}
\def\labelenumi{\arabic{enumi}.}
\item
  \textbf{Residual Vector Quantization (RVQ):} In RVQ schemes,
  quantization is performed hierarchically. Each subsequent layer
  (codebook) quantizes the residual of the approximations from previous
  layers:
  \[r_i = r_{i-1} - \mathbf{q}_i(r_{i-1}), \qquad r_0 = \mathbf{x}\]
  Consequently, the distribution of codes is highly correlated across
  depth (\(P(c_1, c_2, \dots, c_J) \neq \prod_{j=1}^J P(c_j)\)). To
  generate such dependent codes, autoregressive TTS models must either
  use sequential, token-by-token codebook generation (as in VALL-E
  (\protect\hyperlink{ref-wang2023valle}{C. Wang et al. 2023}) or
  Orpheus-TTS), which increases the context sequence length by a factor
  of \(N\) (where \(N\) is the number of codebooks, typically 8) and
  severely slows down generation due to bloated KV caches, or introduce
  specialized internal depth-transformers (as in MusicGen
  (\protect\hyperlink{ref-copet2023musicgen}{Copet et al. 2023}) or
  VALL-E 2 (\protect\hyperlink{ref-xia2024valle2}{Xia et al. 2024})).
  The depth-transformer acts as an auxiliary module modeling the
  conditional joint distribution \(P(c_1, \dots, c_J \mid h)\) along the
  vertical axis (within a single time step). From an inference
  perspective, depth-transformers are incompatible with optimized LLM
  serving engines like vLLM (\protect\hyperlink{ref-kwon2023vllm}{Kwon
  et al. 2023}). Systems like vLLM are designed for a standard
  generation cycle with flat KV-caching and continuous batching.
  Integrating an additional vertical pass through a depth-transformer at
  each decoding step requires custom non-linear operations and dynamic
  computation graph changes, which cannot be implemented without heavily
  modifying the engine's core source code and writing low-level CUDA
  kernels.
\item
  \textbf{GroupFSQ:} Since GroupFSQ channels are orthogonal and
  independent by design, the conditional multi-information (total
  correlation) between channels given the latent state \(h\) is
  negligible (\(I(c_1, \dots, c_{32} \mid h) \approx 0\)). This makes
  factorized parallel sampling across all channels mathematically sound
  and allows us to generate \textbf{the entire audio frame in a single
  autoregressive step}, completely preserving compatibility with
  standard LLM inference and stock vLLM. We selected NeMo NanoCodec
  (\protect\hyperlink{ref-nvidia2025nanocodec}{Casanova et al. 2025}) as
  a SOTA FSQ-based audio tokenizer. External validation of this
  single-pass, frame-by-frame generation without local transformers over
  the codec is provided in the decoder-less mode of Magpie-TTS
  (\protect\hyperlink{ref-nvidia2024magpie}{Neekhara et al. 2024}).
\end{enumerate}

\textbf{Training Complexities with Frame-by-Frame Generation:} We
emphasize that the choice of FSQ was driven \textbf{solely by generation
speed and standard serving engine compatibility}. From a training
perspective, frame-by-frame generation significantly complicates model
convergence compared to sequential codebook-by-codebook models (such as
our earlier iterations, Kani-TTS and Kani-TTS-2, which also used
NanoCodec but with a 4-codebook configuration and sequential sampling).
In a sequential autoregressive step, the transformer predicts a single
discrete token conditioned on previous codebooks, which is a simpler
task with low prediction entropy. In frame-by-frame generation, the
model must predict a vector of 32 independent channels simultaneously: a
highly dense, high-entropy object representing a 1/21.5-second spectral
slice of speech. Modeling this complex joint distribution without
intermediate codebook-level context reduces the model's confidence
during decoding, increases prediction uncertainty, and slows down
backbone convergence during pretraining.

\hypertarget{codec-mixed-radix-unfolding-to-32-heads}{%
\subsubsection{2.3.1. Codec Mixed-Radix Unfolding to 32
Heads}\label{codec-mixed-radix-unfolding-to-32-heads}}

In the original NeMo NanoCodec, the codec output is represented by
\(G=8\) packed tokens, each in the range \(0\dots2015\). In Gepard, we
perform a mixed-radix unfold operation, decomposing each token into 4
independent FSQ channels, yielding \(C = 32\) channels per frame with
cyclically repeating alphabet capacities \(L_k \in \{8,7,6,6\}\)
(detailed mathematical formulas are provided in Appendix A.1).

Instead of a classical architecture with 8 classification heads (where
each head's size matches the packed codebook capacity of 2016, yielding
\(8 \times 2016 = 16,128\) total logits), we transitioned to 32
independent tiny heads with dimensions \(L_k \in \{8,7,6,6\}\) (totaling
\(8 \times 8 + 8 \times 7 + 16 \times 6 = 216\) logits) for the
following reasons: 1. \textbf{Inference Speed:} This drastically reduces
the projection dimensionality of the output classification layer and the
computational overhead of logit generation, yielding a clear speedup
during inference (detailed performance measurements will be presented in
future work). 2. \textbf{Bypassing Redundant Operations:} Internally,
the NeMo NanoCodec decoder unpacks the 8 codebook tokens into the same
32 channels using the same mixed-radix system before speech synthesis.
By predicting the 32 channels directly, we align the backbone's output
directly with the internal representation of the codec and bypass the
redundant decompression math step during generation. 3.
\textbf{Preserving Expressiveness:} There was a concern that using 32
independent tiny classifiers instead of 8 joint 2016-class heads would
degrade the model's expressiveness, as the backbone would lose explicit
modeling of intra-group code correlations. However, experiments showed
that the model successfully compensates for this via distributed
attention in the hidden layers: Gepard demonstrates no loss in synthesis
quality or naturalness compared to Kani-TTS, maintaining full expressive
capacity for timbre and prosody.

\hypertarget{audio-interface-from-codec-codes-to-frame-embedding}{%
\subsection{2.4. Audio Interface: From Codec Codes to Frame
Embedding}\label{audio-interface-from-codec-codes-to-frame-embedding}}

The audio interface is the most engineering-heavy part of the model:
here, discrete codec codes are transformed into a continuous vector that
must align in scale with the text embeddings at the input of the stock
backbone. Most pretraining challenges were concentrated here (the
quantitative aspect is detailed in the training section).

The audio input enters the model already unfolded into 32 channels
(unfolding is performed in the data pipeline, §2.3): integer codes
\(c^{(k)} \in \{0,\dots,L_k-1\}\), \(k = 0..31\), with
\(L_k \in \{8,7,6,6\}\) cyclically. The frame embedding is constructed
as follows:

\[
\begin{aligned}
e_k    &= E_k\!\left[c^{(k)}\right], \qquad E_k \in \mathbb{R}^{L_k \times m}, \quad m = 32 \\
u      &= \bigl[\,e_0 \,;\, e_1 \,;\, \dots \,;\, e_{31}\,\bigr] \in \mathbb{R}^{32 m = 1024} \\
\hat z &= \mathrm{LN}_{\text{aff-free}}\!\bigl(W_2\,\mathrm{GELU}(W_1 u + b_1) + b_2\bigr), \qquad W_1, W_2 \in \mathbb{R}^{1024 \times 1024} \\
x_\text{audio} &= s_\text{audio}\cdot \hat z \in \mathbb{R}^{d}, \qquad s_\text{audio} \leftarrow \operatorname{std}\!\bigl(\text{embed\_tokens}\bigr).
\end{aligned}
\]

Specifically, we perform a channel-wise lookup from 32 tables \(E_k\),
concatenate them into a 1024-dimensional vector, pass it through a
two-layer GELU-MLP, apply an affine-free LayerNorm, and scale by
\(s_\text{audio}\). The resulting frame \(x_\text{audio}\) is
concatenated into the sequence (§2.1) and passed to the backbone, which
applies RMSNorm to the input in its very first layer. Four design
decisions and their justifications:

\begin{enumerate}
\def\labelenumi{\arabic{enumi}.}
\tightlist
\item
  \textbf{MLP instead of sum or average:} Early iterations averaged the
  32 lookups. The average (or sum) is additive (i.e., a linear function
  of channels) and cannot model joint codebook interactions. The
  two-layer GELU-MLP introduces non-linearity, allowing the frame to
  encode the joint structure of the 32 quantizers. Although channels are
  independent at the codec level (§2.3), their embedding into a shared
  vector benefits from non-linear mixing.
\item
  \textbf{Direct path without scale barrier:} An earlier design placed a
  \texttt{Linear\ →\ RMSNorm\ →\ ×0.02} sequence between concatenation
  and the backbone. The \(0.02\) multiplier combined with normalization
  created a gradient barrier of order \(\approx 2400\): audio embeddings
  received gradients thousands of times smaller than text and barely
  trained. Removing this barrier (direct MLP without manual scaling)
  increased the audio gradient norm by 10--20 times.
\item
  \textbf{Affine-free LayerNorm:} The input RMSNorm of the backbone
  discards vector magnitude, making the scale of the frame embedding a
  free direction (not penalized by the loss). Without external
  constraints, the scale of the frame embedding drifted continuously,
  pushing the MLP pre-activations into the saturation region of GELU,
  where it degenerates into a linear function. A LayerNorm without
  learnable parameters (\(\gamma, \beta\)) removes this degree of
  freedom, fixing the norm of \(\hat z\) and preserving the non-linear
  expressiveness of the projection.
\item
  \textbf{Scale alignment with text via \(s_\text{audio}\):} After
  LayerNorm, the output has a unit scale; the non-learnable buffer
  \(s_\text{audio}\) scales it to the standard deviation of text
  embeddings. Although the backbone's RMSNorm will eventually discard
  the scale, this matching keeps the frame in-distribution for
  diagnostics and ensures
  \(\lVert x_\text{audio}\rVert \approx \lVert x_\text{text}\rVert\)
  from the first optimization step, avoiding manual tuning.
  Conceptually, this is manual cross-modality scale alignment
  (\protect\hyperlink{ref-wang2019multimodal}{Z. Wang et al. 2019}),
  related to the Maximal Update Parametrization (\(\mu\)P) framework
  (\protect\hyperlink{ref-yang2022mup}{Yang et al. 2022}).
\end{enumerate}

\hypertarget{output-heads-and-loss-function}{%
\subsection{2.5. Output Heads and Loss
Function}\label{output-heads-and-loss-function}}

The output heads are applied only to the audio region of the hidden
states. Since labels are constructed for the
\texttt{{[}text\ \textbar{}\ audio{]}} region (without the prefix), the
\(K\)-position prefix slice is discarded before the heads; otherwise,
the causal alignment of the loss would be violated.

\begin{itemize}
\tightlist
\item
  \textbf{32 codebook heads:} One linear layer
  \(\mathbb{R}^{d}\to\mathbb{R}^{L_k}\) per channel, with its respective
  alphabet size \(L_k\). The loss is cross-entropy with a causal shift.
\item
  \textbf{1 stop head:} \(\mathbb{R}^{d}\to\mathbb{R}\), binary sigmoid,
  predicting the end of speech (the terminal ``phantom'' audio frame is
  labeled with \(\text{stop}=1\)).
\end{itemize}

The core loss is:

\[
\mathcal{L} = \underbrace{\sum_{k=0}^{31} \mathrm{CE}\bigl(\text{logits}_k,\, \text{labels}_k\bigr)}_{\text{32 codebook heads}}
\;+\; w_\text{stop}\cdot \mathrm{BCE}_{\text{pos\_weight}}\bigl(\text{stop\_logits},\, \text{stop\_labels}\bigr),
\]

with \(w_\text{stop} = 2\). The class \(\text{stop}=1\) is rare (approx.
1 frame out of 150), so BCE is calculated with
\texttt{pos\_weight\ =\ 25}: without reweighting, the loss collapses to
``always 0,'' and the stop threshold is never crossed at inference. The
cross-entropy of each head is protected from the degenerate case where
all labels after shifting are \texttt{-100} within a microbatch: instead
of \texttt{mean}-CE (which would yield \(0/0 = \text{NaN}\)), we compute
\(\text{logits}\cdot 0\), keeping the head in the graph with zero
gradient.

These two terms form the core training objective. When voice cloning is
active, two compressor regularizers are added:

\[
\mathcal{L}_\text{total} = \sum_{k=0}^{31} \mathrm{CE}_k \;+\; w_\text{stop}\,\mathrm{BCE}_\text{stop}
\;+\; \underbrace{\mathcal{L}_\text{div} \;+\; \mathcal{L}_\text{supcon}}_{\text{VC compressor, §2.6}}.
\]

The terms \(\mathcal{L}_\text{div}\) (diversity) and
\(\mathcal{L}_\text{supcon}\) (supervised contrastive) are described in
§2.6. During the fine-tuning and DPO stages, the compressor is frozen
and these regularizers do not participate.

\hypertarget{voice-cloning-q-former-prefix}{%
\subsection{2.6. Voice Cloning: Q-Former
Prefix}\label{voice-cloning-q-former-prefix}}

Voice cloning is an optional feature. When disabled, all cloning
branches are bypassed and the forward pass matches the pre-VC behavior.
Voice identity is extracted from a reference audio clip and prepended as
a prefix before the text, aligning with the principle in §1.2: the
prefix is computed once in prefill and is absent from the decode loop.

\hypertarget{compressor-q-former}{%
\subsubsection{2.6.1. Compressor (Q-Former)}\label{compressor-q-former}}

\begin{figure}
\centering
\includegraphics[width=0.88\textwidth,height=\textheight]{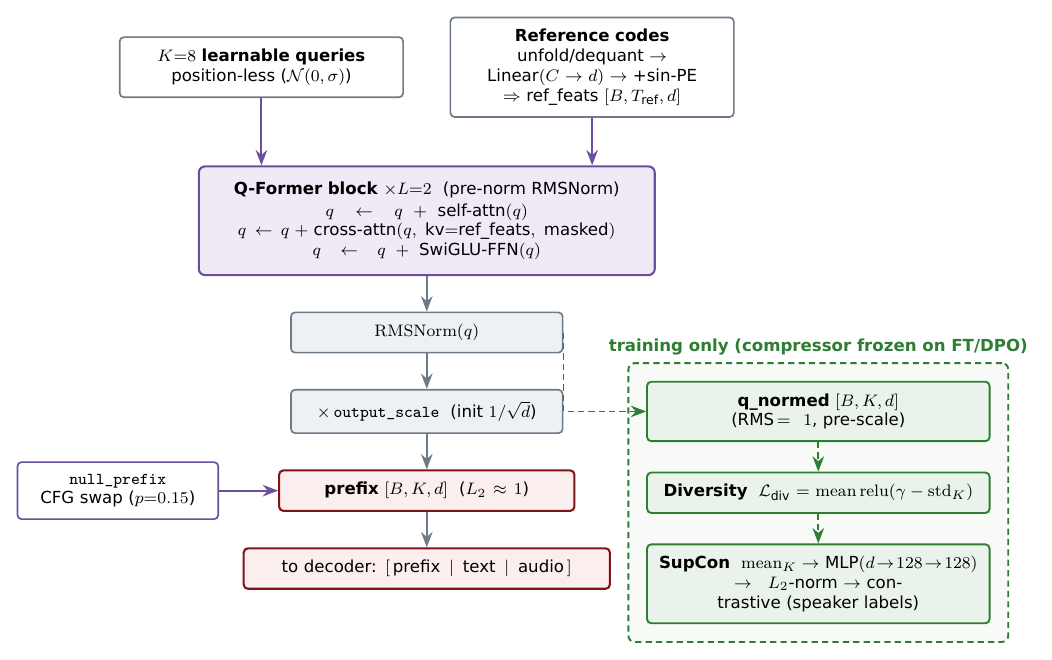}
\caption{Q-Former voice-cloning compressor (training-time view).
Position-less queries and reference features (with sinusoidal PE) pass
through \(L{=}2\) pre-norm blocks; \(\mathrm{RMSNorm}(q)\) feeds the
scaled \texttt{prefix} (with \texttt{null\_prefix} CFG swap) into the
decoder, while \texttt{q\_normed} feeds the diversity and SupCon
regularizers, which are active only during compressor training. Verified
against \texttt{utils/ref\_compressor.py} and
\texttt{utils/supcon.py}.\label{fig:refcompressor}}
\end{figure}

A reference codec token stack of variable length is compressed into
\(K = 8\) speaker tokens. Reference codes undergo unfold/dequantize,
linear projection to \(d\), and sinusoidal positional encoding (applied
only to reference features, queries are position-less). Then, \(K\)
learnable query tokens are passed through \(L = 2\) Q-Former blocks
(\protect\hyperlink{ref-li2023blip2}{Li et al. 2023}) (self-attention →
masked cross-attention to reference features → SwiGLU FFN, pre-norm
RMSNorm, 8 heads), a structure related to architectures like Perceiver
IO (\protect\hyperlink{ref-jaegle2021perceiverio}{Jaegle et al. 2021})
and Flamingo (\protect\hyperlink{ref-alayrac2022flamingo}{Alayrac et al.
2022}). The output is scaled as
\(\text{output\_scale}\cdot\mathrm{RMSNorm}(q)\) with initialization
\(\text{output\_scale} = 1/\sqrt{d}\), ensuring the starting
\(L_2\)-norm of the prefix is \(\approx 1\) and does not overpower text
and audio embeddings. The query size \(K = 8\) is chosen as a bottleneck
to prevent the compressor from copying the reference clip verbatim
(``copy-paste'' leakage). The compressor parameters and the
\texttt{null\_prefix} are trained with a reduced learning rate
multiplier (\(\times 0.1\)). The compressor outputs two tensors:
\texttt{prefix\_raw} (consumed by the decoder) and \texttt{q\_normed}
(\(\text{RMS} = 1\) per token), which is used to calculate the
regularizers (§2.6.3). A detailed view of the compressor and its
training-time regularizer taps is given in Figure
\ref{fig:refcompressor} (Appendix A.4).

\hypertarget{null_prefix-and-cfg-dropout}{%
\subsubsection{\texorpdfstring{2.6.2. \texttt{null\_prefix} and
CFG-Dropout}{2.6.2. null\_prefix and CFG-Dropout}}\label{null_prefix-and-cfg-dropout}}

The learnable parameter \texttt{null\_prefix}
\(\in \mathbb{R}^{K\times d}\) (initialized with standard deviation
\(0.02\)) defines the unconditional path required for Classifier-Free
Guidance at inference (§6.1). It replaces the real prefix on a
per-sample basis via two independent triggers combined with an OR:

\begin{enumerate}
\def\labelenumi{\arabic{enumi}.}
\tightlist
\item
  Stochastic CFG-dropout with probability
  \texttt{cfg\_dropout\_prob\ =\ 0.15} (active only during training).
\item
  Forced substitution for lines with null-sentinel speakers:
  low-frequency speakers (\textless{}
  \texttt{min\_clips\_per\_speaker\ =\ 3}) and speaker-less sources are
  not discarded, but trained unconditionally
  (\texttt{singleton\_policy:\ null\_prefix}).
\end{enumerate}

The total fraction of unconditional exposure is the structural fraction
of sentinel rows (\(\approx 3.05\%\), §4.1.1) plus stochastic dropout
over the remaining samples. Both fractions are logged separately to
distinguish drift from sentinel data from the CFG-dropout itself. This
guarantees that even the dense ( \(\ge K\) clips) bucket sees the
unconditional path, without which inference CFG cannot function.

\hypertarget{compressor-regularizers-representation-leakage-and-supcon}{%
\subsubsection{2.6.3. Compressor Regularizers: Representation Leakage
and
SupCon}\label{compressor-regularizers-representation-leakage-and-supcon}}

Both regularizers are computed on the normalized representation
\(q_\text{normed}\) (with \(\text{RMS} = 1\) per token) before
CFG-dropout. This ensures that regularization acts on the original
output of the compressor rather than the substituted
\texttt{null\_prefix}. The regularizers are introduced into training via
a curriculum (linear ramp-up after a warm-up phase), which excludes
initial noise until the basic phonetic structure in the decoder
stabilizes.

Using quantized representations (such as FSQ) in zero-shot voice cloning
carries the risk of \textbf{representation leakage}. Since the decoder
aims to minimize reconstruction error, the compressor receives gradient
incentives to encode the detailed spectral portrait of the reference
clip (acoustic footprint) rather than an invariant speaker timbre. In
the degenerate case, the compressor resolves reconstruction by simply
copying the reference (``copy-paste'' strategy), completely ignoring
text conditioning. Since both strategies mathematically yield identical
reconstruction losses, standard training cannot separate these modes.

To overcome this issue, a regularization system was developed:

\begin{enumerate}
\def\labelenumi{\arabic{enumi}.}
\item
  \textbf{Diversity Loss (hinge-variance):} Prevents the collapse of the
  \(K\) latent queries into a single vector, forcing the model to
  utilize the allocated prefix capacity:
  \[\mathcal{L}_\text{div} = \frac{1}{K}\sum_{j=1}^K \mathrm{relu}\bigl(\gamma - \operatorname{std}\nolimits_K(q_\text{normed})\bigr), \qquad \gamma = 0.5\]
  where the standard deviation is computed over the dimension \(K\), and
  \(\gamma\) is a variance threshold parameter.
\item
  \textbf{Supervised Contrastive Loss (SupCon):} Acts as a key semantic
  filter (\protect\hyperlink{ref-khosla2020supcon}{Khosla et al. 2020}),
  related to joint-embedding techniques like VICReg
  (\protect\hyperlink{ref-bardes2021vicreg}{Bardes, Ponce, and LeCun
  2021}). It forces the compressor's representation to be invariant to
  the text content of the clip and sensitive only to speaker identity.
  To achieve this, \(q_\text{normed}\) is averaged over the \(K\) tokens
  into a single vector \(z_i\), passed through a projection head
  (2-layer MLP, \(\approx 260\text{K}\) parameters), and
  \(L_2\)-normalized. The loss formula is:
  \[\mathcal{L}_\text{supcon} = \frac{1}{|A|}\sum_{i\in A} \frac{-1}{|P(i)|}\sum_{p\in P(i)} \log \frac{\exp(\langle z_i, z_p \rangle /\tau)}{\sum_{a\in V, a\neq i}\exp(\langle z_i, z_a \rangle /\tau)}\]
  where \(V\) is the set of all valid (active, non-sentinel) samples in
  the batch, \(P(i) \subseteq V \setminus \{i\}\) is the subset of
  samples belonging to the same speaker as anchor \(i\), and
  \(A \subseteq V\) is the subset of active anchors with at least one
  positive example in the batch (\(|P(i)| \geq 1\)). The temperature
  parameter is \(\tau = 0.1\).

  During training, the batch is structured by a specialized sampler
  using a \(P\cdot K + M\) scheme (where \(P=16\) speakers, \(K=3\)
  clips per speaker, and \(M=16\) background null-sentinel clips).
  Background clips are excluded from the anchors \(A\) and positive
  pairs \(P(i)\), but act as negative examples in the denominator,
  expanding the contrastive subspace. Negative examples are further
  scaled using cross-rank grouping without gradient retention for remote
  hosts, increasing the effective batch size to
  \(W_\text{size} \times P K\).
\end{enumerate}

\hypertarget{freeze-during-fine-tuning-and-observations-on-transfer}{%
\subsubsection{2.6.4. Freeze during Fine-Tuning and Observations on
Transfer}\label{freeze-during-fine-tuning-and-observations-on-transfer}}

During the fine-tuning and DPO stages, the compressor is frozen (it was
trained only during pretraining). The prefix remains but is inert, so
regularizer losses and specialized batch sampling are disabled.

\emph{Observation: Timbre-only transfer.} Qualitatively (not yet
measured formally, but verified on the demo page), the compressor
transfers timbre from the reference, but not accent or language. If the
prompt audio is in Russian and the generated text is in English, the
output sounds like the reference speaker but speaks fluent English
without a Russian accent. The same reference works for Dutch, Spanish,
and other languages. This aligns with the mechanics of SupCon (§2.6.3):
the \(K = 8\) speaker tokens carry a low-dimensional representation that
the contrastive loss explicitly forced to be content-invariant. Since
the reference phonetics do not leak into this bottleneck, only timbre
remains. Unlike many zero-shot VC systems that transfer the speaker's
native accent, this is a potential differentiator and a candidate for a
cross-lingual evaluation.

\begin{center}\rule{0.5\linewidth}{0.5pt}\end{center}

\hypertarget{engineering-discoveries-verified-by-experiment}{%
\section{3. Engineering Discoveries Verified by
Experiment}\label{engineering-discoveries-verified-by-experiment}}

This section is dedicated to design choices implemented in response to
specific challenges and validated by measurements. They are highlighted
because they are reusable: building a codec-audio interface on a
pretrained LLM backbone in a decoder-only TTS introduces similar
difficulties, and the diagnostic tools described below are applicable
beyond Gepard.

The empirical data in this section is taken from the main pretraining
run: 4\(\times\) RTX 6000, FSDP2, \(\approx 7\) epochs out of 8 planned,
with an intermediate checkpoint at \(\approx 117\text{k}\) steps.
Diagnostic metrics were logged throughout training.

\hypertarget{modality-scale-alignment-framework}{%
\subsection{3.1. Modality Scale Alignment
Framework}\label{modality-scale-alignment-framework}}

Decoder-only TTS connects two fundamentally different entities:
pretrained text representations (high-level, requiring slow adaptation)
and audio representations learned from scratch (low-level, with high
learning dynamics). Direct naive concatenation of these modalities
triggers two main problems: gradient imbalance (gradient starvation of
one of the modalities) and unbounded drift of latent scales. To address
these issues, we developed a modality scale alignment framework
consisting of four components:

\begin{enumerate}
\def\labelenumi{\arabic{enumi}.}
\tightlist
\item
  \textbf{Elimination of Gradient Barriers:} In early protocols, a
  \texttt{Linear\ -\textgreater{}\ RMSNorm\ -\textgreater{}\ x0.02}
  sequence was placed between the audio embedding concatenation and the
  backbone input. The presence of the rigid scaling factor of \(0.02\)
  led to the gradient norm for the audio interface being
  \(\approx 2400\) times smaller than the text embedding gradient norm.
  The audio components did not train. Removing this artificial
  multiplier and switching to direct MLP coupling increased the gradient
  norm of the audio interface by 10--20 times, balancing the learning
  dynamics (see Appendix B.2 for the detailed backpropagation chain-rule
  analysis).
\item
  \textbf{Fixing the Latent Norm (Affine-free LayerNorm):} Since the
  first layer of the backbone applies RMSNorm to the combined sequence,
  the absolute magnitude of the vectors is a free direction (not
  penalized by the loss). Without external constraints, the scale of the
  frame embedding increased monotonically during optimization, pushing
  the MLP pre-activations into the saturation region of GELU, where it
  degenerates into a linear function. Using LayerNorm without learnable
  scale and shift parameters (\(\gamma, \beta\)) removes this degree of
  freedom, fixing the norm of \(\hat{z}\) and preserving the non-linear
  expressiveness of the projection.
\item
  \textbf{Weight Parametrization via \(\mu\)P Principles:} To align
  scales from the very first optimization step, the initialization
  variance of the audio tables \(E_k\) is set to \(1.0\), and the
  weights of the MLP linear layers are initialized with a variance
  inversely proportional to the input size (\(in^{-1/2}\), with biases
  initialized to zero). The output scale is aligned with the variance of
  the pretrained text embedding via a fixed non-learnable coefficient:
  \[s_\text{audio} \leftarrow \operatorname{std}(\text{embed\_tokens})\]
  This manually aligns the scales of the two modalities, preventing
  representations from drifting apart at the start (a numerical
  breakdown of the initial \(60\times\) scale mismatch is provided in
  Appendix B.1).
\item
  \textbf{Split Learning Rates:} To prevent catastrophic forgetting and
  preserve the structure of the pretrained text space, the learning rate
  for the text embedding is scaled by a factor of
  \(\eta_{\text{text}} = 0.2\), and for the audio interface by
  \(\eta_{\text{audio}} = 0.5\) relative to the base optimizer step
  (\protect\hyperlink{ref-kumar2022lpft}{A. Kumar et al. 2022}).
\end{enumerate}

\begin{figure}
\centering
\includegraphics[width=1\textwidth,height=\textheight]{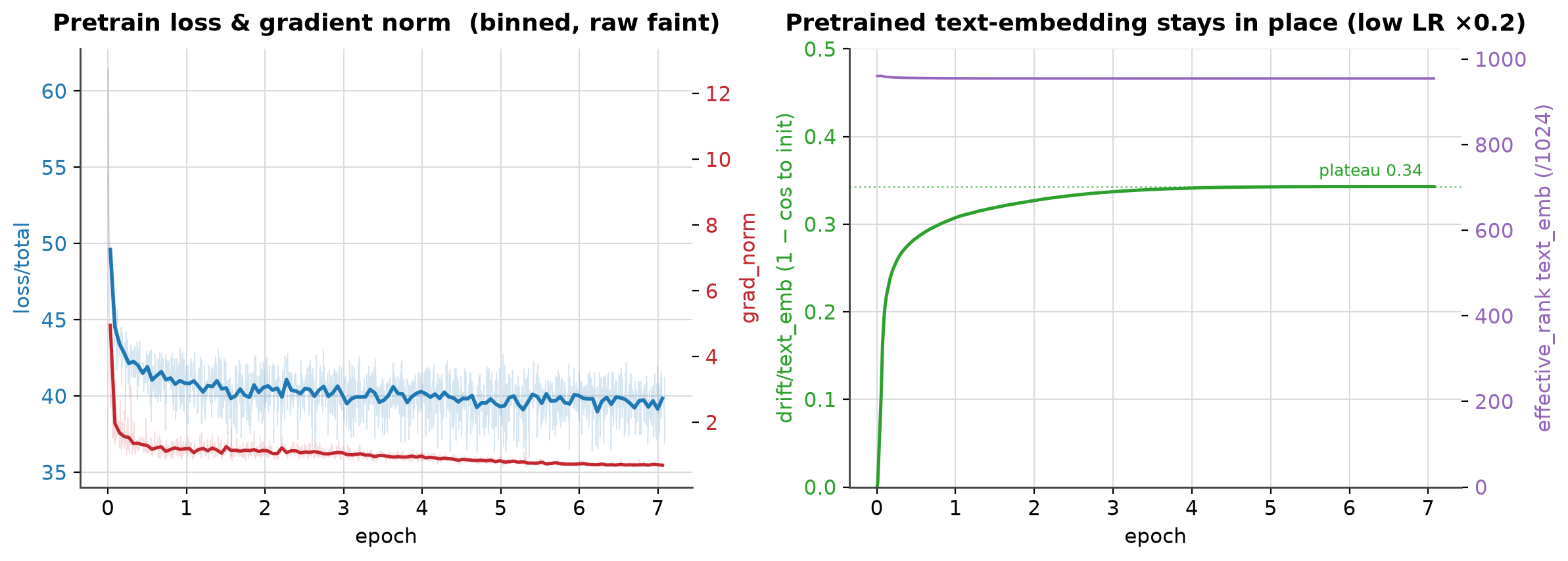}
\caption{Diagnostics of modality alignment during pretraining. Left:
\texttt{loss/total} and gradient norm (binned). Right: text embedding
drift flattens at \(\approx 0.34\), while the effective rank stays
around \(\approx 955/1024\), indicating adaptation without losing the
pretrained structure.\label{fig:scaling-curves}}
\end{figure}

The quantitative effect of the framework is confirmed by pretraining
diagnostics. The text embedding adapts without structural collapse: the
drift norm relative to initialization stabilizes at \(\approx 0.34\) by
the 3rd epoch (cosine similarity to the starting point is \(0.948\)),
and the effective rank of the text representation matrix maintains a
value of \(\approx 955\) out of \(1024\) throughout optimization (Figure
\ref{fig:scaling-curves}, right panel). Importantly, while the text and
audio representations remain orthogonal in weight space, hidden-state
monitoring reveals that they successfully align in the deeper layers of
the backbone (as discussed in Appendix B.3).

The gradient regime remains stable: after opening the gradient path, the
total loss converges without singularities, and the gradient norm
smoothly decreases to a baseline level of \(\approx 1.5\) (Figure
\ref{fig:scaling-curves}, left panel). The loss flattening on the
plateau is due to the computational budget limit rather than network
capacity saturation.

\emph{Effective rank as a diagnostic indicator.} We use the
entropy-based effective rank
(\protect\hyperlink{ref-roy2007effective}{Roy and Vetterli 2007}) of the
weight matrix \(W\) (computed via its singular values \(\sigma_i\)) as a
diagnostic indicator of degradation:

\[
\operatorname{eff\_rank}(W) = \exp\!\left(-\sum_{i} p_i \log p_i\right),
\qquad p_i = \frac{\sigma_i}{\sum_j \sigma_j}.
\]

\begin{figure}
\centering
\includegraphics[width=1\textwidth,height=\textheight]{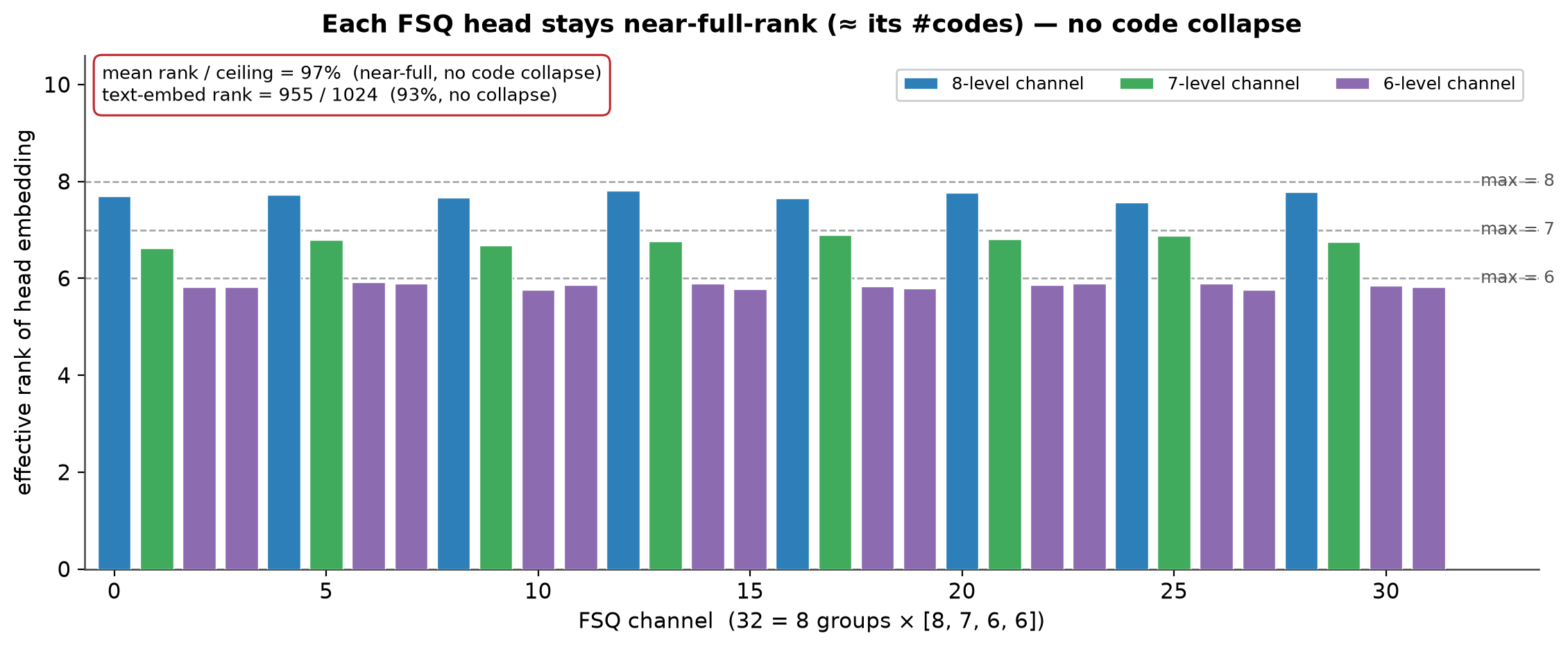}
\caption{Effective rank of the 32 audio lookup tables during
pretraining. The dashed line is the structural ceiling (the number of
FSQ codes of the channel). All heads remain close to the ceiling,
indicating full capacity utilization and no intra-head
collapse.\label{fig:effrank}}
\end{figure}

Here, interpretation accuracy is crucial. The audio channel table has a
shape of \([L_k \times 32]\), where \(L_k \in \{8,7,6,6\}\) is the
number of FSQ codes; thus, its effective rank is structurally bounded
from above by the number of codes \(L_k\). The observed rank is
\(\approx 96\text{--}97\%\) of this ceiling across all 32 channels
(Figure \ref{fig:effrank}): each table utilizes its full capacity, and
codes remain linearly distinguishable. This indicates the absence of
collapse rather than emergent compression to a lower dimension (rank
cannot exceed the number of rows). The same indicator for the text
embedding (\(955/1024\)) reads similarly: high rank, no collapse.

\hypertarget{groupfsq-and-factorized-sampling-without-depth-transformer}{%
\subsection{3.2. GroupFSQ and Factorized Sampling without
Depth-Transformer}\label{groupfsq-and-factorized-sampling-without-depth-transformer}}

A recurring question in decoder-only TTS with multi-codebook codecs is
whether a depth- or local-transformer is required over the codebooks to
model their joint distribution. For GroupFSQ, the answer is negative,
which can be shown from an information-theoretic perspective.

Let \(h\) be the latent state of a frame, and \(c_1,\dots,c_{32}\) be
its 32 channels. A factorized (parallel across heads) head pays the sum
of conditional entropies, whereas the true joint distribution costs the
joint conditional entropy:

\[
\text{Factorized: } \sum_{i=1}^{32} H(c_i \mid h),
\qquad
\text{True: } H(c_1,\dots,c_{32} \mid h).
\]

The gap between them is the multi-information (total correlation):

\[
I \;=\; \sum_{i=1}^{32} H(c_i \mid h) \;-\; H(c_1,\dots,c_{32} \mid h) \;\ge\; 0,
\]

and it is precisely \(I > 0\) that causes factorized sampling errors.
For GroupFSQ, the channels are independent by design (this is not
residual VQ; there is no dependency where ``channel \(i\) encodes the
residual of channel \(i-1\)''), meaning \(I \approx 0\), and
factorization is correct. A depth-transformer is unnecessary and,
moreover, would introduce a custom operation into the decode loop,
conflicting with the principle in §1.2.

This analytical argument is supported by external evidence: the
Magpie-TTS model (\protect\hyperlink{ref-nvidia2024magpie}{Neekhara et
al. 2024}), which uses an optional local transformer over codebooks,
synthesizes successfully without it (as verified by our tests on
Magpie). A full ablation study of depth-head on/off on our model was not
conducted and remains future work.

\hypertarget{stop-head-as-a-bernoulli-predictor-class-imbalance-and-saturation}{%
\subsection{3.3. Stop Head as a Bernoulli Predictor: Class Imbalance and
Saturation}\label{stop-head-as-a-bernoulli-predictor-class-imbalance-and-saturation}}

Gepard predicts the end of speech not by an EOS token in the vocabulary,
but by a separate binary stop head (§2.5). Generation is best described
as a stochastic process with two sequential decisions at each step
\(t\):

\begin{enumerate}
\def\labelenumi{\arabic{enumi}.}
\tightlist
\item
  A binary decision of whether to end speech (\(s_t = 1\)) or continue
  speaking (\(s_t = 0\)), modeled as a Bernoulli distribution:
  \[s_t \sim \mathrm{Bernoulli}(p_{\text{stop}, t})\]
\item
  Audio code selection: independent sampling of \(C = 32\) channels from
  categorical distributions:
  \[y_t^{(c)} \sim \mathrm{Categorical}(\mathbf{p}_t^{(c)}), \qquad c = 1..C\]
\end{enumerate}

The total log-likelihood of a generated trajectory \(y\) of length \(T\)
frames (from \(t = 0\) to \(T-1\)) given input \(x\) is:

\[\log\pi(y \mid x) = \sum_{t=0}^{T-1} \sum_{c=1}^{C} \log p_c\bigl(y_t^{(c)}\bigr) + \sum_{t=0}^{T-2}\log(1 - p_{\text{stop},t}) + \mathbb{I}(\text{not truncated}) \log p_{\text{stop},T-1}\]

where \(\mathbb{I}(\text{not truncated})\) is the indicator function
taking the value 1 if the generation finished via natural stop
(predicting \(s_{T-1} = 1\)), and 0 if the trajectory was truncated upon
reaching the maximum frame limit. The Bernoulli terms
\(\log(1-p_{\text{stop},t})\) and the final term
\(\log p_{\text{stop},T-1}\) are exact mathematical equivalents of the
EOS token probability in classical autoregressive language models, but
separated into an independent projection branch.

Class imbalance distorts calibration. Since the stop event (\(s_t = 1\))
occurs only once per trajectory (approx. 1 frame out of 150), a standard
binary cross-entropy (BCE) loss on pretraining converges to a local
minimum with average \(p_{\text{stop}, t} \approx 0.05\). At inference,
such a model never crosses the decision threshold of 0.5. To compensate
for this imbalance, BCE optimization is performed with a positive class
weight \(w_{\text{pos}} = 25.0\) (§2.5).

During training, the stop head saturates rapidly, approaching a delta
function. The empirical probability distribution of
\(p_{\text{stop}, t}\) displays a binary nature: on \(99.7\%\) of frames
outside the stop zone, the probability lies within
\(0.0001 \dots 0.004\), after which it jumps to \(1.0\) at the terminal
frame. This observation leads to two important conclusions:

\begin{enumerate}
\def\labelenumi{\arabic{enumi}.}
\tightlist
\item
  \textbf{Inference heuristics are ineffective:} Attempts to sample the
  stochastic stop, vary the decision threshold, or apply temperature
  scaling to stop logits do not make sense because there is no
  transition zone (\(0.1 \dots 0.5\)) in the probability distribution.
  Any anomalies in stop behavior must be corrected during training.
\item
  \textbf{Critical role of Bernoulli terms in credit assignment:} During
  DPO optimization (Section 5), the Bernoulli components must be
  included in the trajectory log-likelihood \(\log\pi\); otherwise,
  preference gradients will not act on the stop decision. Due to the
  pronounced saturation of the head (where logit values go to
  \(\pm\infty\)), the gradient of the sigmoid activation function at the
  extremes approaches zero. To prevent gradient vanishing, we clip the
  predicted probability with a floor \(p_{\text{floor}} = 10^{-4}\),
  i.e.,
  \(p_{\text{stop}} \leftarrow \max(p_{\text{stop}}, p_{\text{floor}})\).
\end{enumerate}

Finally, diagnostics showed that the stop head is not ``blind'': in a
subset of failures, it outputs 1.0 a few seconds late (late-stop mode),
meaning it triggers successfully on self-generated states when the
backbone exits the ``I am speaking'' mode. This localizes the root of
the runaway on short inputs to the backbone rather than the stop head, a
conclusion finalized in the chapter on the short register (Section 5)
and supported by quantitative variance diagnostics (Appendix B.4), with
structural parallels to VALL-T (\protect\hyperlink{ref-du2024vallt}{Du
et al. 2024}).

\begin{center}\rule{0.5\linewidth}{0.5pt}\end{center}

\hypertarget{data-and-training}{%
\section{4. Data and Training}\label{data-and-training}}

Training consists of pretraining and two fine-tuning stages (SFT,
followed by DPO). Data and pretraining are described in §4.1--4.2; the
fine-tuning strategy and SFT stage are covered in §4.3. The DPO stage
serves to address the short register failure mode and is expanded in
Section 5; here, only its role in the pipeline is outlined. Training
stability (the NaN incident and guards) is detailed in §4.4.

\hypertarget{data}{%
\subsection{4.1. Data}\label{data}}

\hypertarget{composition-and-scale}{%
\subsubsection{4.1.1. Composition and
Scale}\label{composition-and-scale}}

The training set is a multilingual mix of datasets tokenized by
NanoCodec (§2.3). The full pretraining mix contains 27,623,833 samples
and 68,833 hours of audio (247.8 M s) from 19 sources, with 1,675,752
speakers after filtering with \texttt{min\_clips\_per\_speaker}. The
frame rate is 21.5 Hz. Clip duration: mean 8.97 s, median 7.44 s, p95
19.2 s, max 40 s; the distribution is concentrated in the 3--10 s range
(\(\approx 64\%\)), and the 20--40 s tail constitutes \(\approx 3.8\%\).
Text length (clean tokens, excluding service SOT/EOT/SOS tokens): mean
32.7, median 27, p95 75; the 10--50 token range covers \(\approx 78\%\)
of the corpus. The fraction of rows routed through \texttt{null\_prefix}
(speaker-less and singletons) is
\texttt{pct\_structural\_null\_exposure} \(\approx 3.05\%\); the number
of SupCon-eligible rows is \(\approx 26.8\text{ M}\) with
\(\approx 1.04\text{ M}\) speakers.

\hypertarget{languages}{%
\subsubsection{4.1.2. Languages}\label{languages}}

\begin{longtable}[]{@{}
  >{\raggedright\arraybackslash}p{(\columnwidth - 4\tabcolsep) * \real{0.2989}}
  >{\raggedright\arraybackslash}p{(\columnwidth - 4\tabcolsep) * \real{0.4713}}
  >{\raggedright\arraybackslash}p{(\columnwidth - 4\tabcolsep) * \real{0.2299}}@{}}
\toprule\noalign{}
\begin{minipage}[b]{\linewidth}\raggedright
Language / Variant
\end{minipage} & \begin{minipage}[b]{\linewidth}\raggedright
Source Description
\end{minipage} & \begin{minipage}[b]{\linewidth}\raggedright
\(\approx\) Rows
\end{minipage} \\
\midrule\noalign{}
\endhead
\bottomrule\noalign{}
\endlastfoot
English (+ Boston, Glasgow, NY, Oakland, Scouse accents) & Podcast
corpus, dialectal English, Taste Dump (17.8 M), Emolia (filtered) (5.4
M) & Dominates (\(\approx 85\%\) of the mix) \\
Spanish (Iberian) & Conversational Spanish, synthetic numeral corpus
(EN/ES) & \(\approx\) 0.4 M \\
Spanish (Mexican) & Podcast corpus, conversational Spanish (MX) &
\(\approx\) 0.44 M \\
Portuguese (Brazilian) & Podcast corpus, synthetic numeral corpus
(NL/PT-BR) & \(\approx\) 0.17 M \\
Dutch & Podcast corpus, synthetic numeral corpus (NL/PT-BR) &
\(\approx\) 0.37 M \\
Kyrgyz & Audiobooks and speech in Kyrgyz (various speakers) &
\(\approx\) 1.15 M \\
\end{longtable}

Additionally, the model was trained to pronounce spoken numbers using
specialized synthetic numeral corpora totaling about 1.4 M rows,
covering number pronunciation in EN/ES/NL/PT-BR.

\hypertarget{data-provenance-and-balance}{%
\subsubsection{4.1.3. Data Provenance and
Balance}\label{data-provenance-and-balance}}

English dominates because high-quality open TTS corpora are
predominantly English-centric: Taste Dump (\(\approx 64\%\) of the mix)
and the filtered version of Emolia (\(\approx 20\%\)) are open sources
with a prevalence of English and emotional speech. Regional accents of
English and other languages were collected using specialized tools for
processing open audio recordings and podcasts. The non-English coverage
demonstrates the scalability of the data preparation pipeline, rather
than volume parity with English.

The language imbalance has a direct consequence: the backbone is
multilingual, but the audio interface was trained predominantly on
English data. Consequently, English is the model's strong suit, while
quality in other languages is limited by data volume and is expected to
grow during subsequent fine-tuning (§1.5). The use of the filtered
Emolia corpus and English-language podcasts during the early pretraining
phase amplified this imbalance, leading to localized degradation of the
audio interface outside the English domain.

\hypertarget{policies-and-processing}{%
\subsubsection{4.1.4. Policies and
Processing}\label{policies-and-processing}}

\begin{itemize}
\tightlist
\item
  \texttt{singleton\_policy:\ null\_prefix},
  \texttt{min\_clips\_per\_speaker:\ 3}: Speakers with fewer than 3
  clips and speaker-less sources are not discarded, but routed through
  \texttt{null\_prefix}. This also ensures the SupCon composition of the
  batch, guaranteeing \(K\) positives in the bucket.
\item
  \texttt{max\_duration\_sec:\ 40}, \texttt{add\_speaker\_id:\ true},
  \texttt{speaker\_statistics:\ true}.
\end{itemize}

In the data pipeline, codec codes are unfolded into 32 channels
(mixed-radix unfolding, §2.3), so audio arrives at the model already
unfolded.

\hypertarget{text-repetition}{%
\subsubsection{4.1.5. Text Repetition}\label{text-repetition}}

Text repetition (\texttt{text\_repetition}, implemented in the text
repetition module) is a key fine-tuning augmentation against collapse on
short inputs, inspired by the VoiceStar approach
(\protect\hyperlink{ref-voicestar2025}{Peng et al. 2025}). On short
text, the prefix \(K = 8\) dominates over 1--2 text tokens, and the
model fails to lock onto the speech variety. The augmentation repeats
the text block until the text region reaches the target token budget:

\begin{verbatim}
[ (SOT text EOT) × (R-1) | SOT text EOT SOS | audio ]
\end{verbatim}

Only the final (canonical) copy carries the SOS service token, so it
alone triggers audio rendering; the context copies are masked with
\texttt{-100} in the labels to exclude supervision and double voicing.
Parameters: \texttt{target\_text\_tokens:\ 16},
\texttt{apply\_below:\ 13} (repetition only for texts shorter than 13
tokens), \texttt{max\_repeats:\ 8}, and
\texttt{mixed\_keep\_prob:\ 0.25} (a portion of short lines is kept at
\(R=1\) so the model learns that repetition is optional). The same logic
is used during data preparation and at inference; a mismatch leads to a
collapse in WER. The calibration of the target budget
(\texttt{target\ =\ 16}) is derived from failure-mode analysis and is
presented in Section 5.

\hypertarget{pretraining}{%
\subsection{4.2. Pretraining}\label{pretraining}}

The base model is a modified text backbone based on the Qwen3.5
architecture (LinearBlocks removed, leaving only full attention; only
the text embedding and tokenizer were inherited from the original model,
the rest was trained from scratch, §2.2). During TTS model assembly, the
weights of this full-attention backbone are loaded, the output
classification layer is discarded, and the audio tables are initialized
from the distribution of text embeddings.

Pretraining parameter configuration:

\begin{itemize}
\tightlist
\item
  Hardware: 4\(\times\) RTX 6000, FSDP2.
\item
  Effective batch:
  \texttt{per\_device\_train\_batch\_size\ =\ 64\ ×\ gradient\_accumulation\ =\ 2\ ×\ 4\ GPU\ =\ 512}.
\item
  Optimization: LR \(9\cdot10^{-4}\), cosine schedule, warmup 500 steps,
  \texttt{max\_grad\_norm} 3.0, bf16, \texttt{adamw\_torch\_fused},
  gradient checkpointing, weight decay 0, \(\beta = (0.9, 0.999)\).
  Separate learning rate multipliers: audio \(\times 0.5\), text
  embedding \(\times 0.2\) (§3.1).
\item
  Backbone: hidden dimension 1024, 14 layers, 8 attention heads (2 KV
  heads), intermediate dimension 3584, \texttt{head\_dim} 256,
  \texttt{tie\_word\_embeddings\ =\ True},
  \texttt{use\_cache\ =\ False}.
\item
  Voice cloning enabled (Q-Former, \texttt{null\_prefix}, SupCon,
  diversity).
\item
  Duration: started 2026-05-19, runtime \(\approx 90\text{ hours}\); the
  run reached \texttt{train/epoch\ ≈\ 7.09} out of 8 planned at step
  116,930 and crashed near the end of the schedule.
\end{itemize}

Following pretraining, the base model (555.7M parameters) showed
acceptable voice quality and speech intelligibility; \texttt{loss/total}
flattened at \(\approx 40\) due to the compute budget rather than
network capacity limitations (§3.1). Outstanding issues at this stage:

\begin{enumerate}
\def\labelenumi{\arabic{enumi}.}
\tightlist
\item
  The stop mechanism failed in approximately 20\% of cases (leading to
  infinite runaway loops at inference).
\item
  Voice cloning did not generalize to unseen voices: the model tended to
  select a similar voice from the training set. The cause was
  representation leakage: the reference clip and target audio always
  belonged to the same speaker, meaning there was no incentive to
  generalize timbre to new voices.
\item
  Slight robotic tone and omission of individual phonemes.
\end{enumerate}

The primary issue was instability on short inputs (defect 1), the
detailed analysis of which is provided in Section 5.

\hypertarget{fine-tuning-two-stage-strategy}{%
\subsection{4.3. Fine-Tuning: Two-Stage
Strategy}\label{fine-tuning-two-stage-strategy}}

The initial plan to ``freeze the backbone and train only the audio
heads'' was revised: text repetition (§4.1.5) defines a new role for the
backbone that a frozen backbone cannot learn. Therefore, fine-tuning
applies LoRA to the backbone. The strategy is two-stage: SFT, followed
by DPO.

\hypertarget{sft-stage}{%
\subsubsection{4.3.1. SFT Stage}\label{sft-stage}}

The first fine-tuning experiment (LoRA-SFT) yielded a noticeable
improvement in intelligibility (WER decreased by 19--64\% across
speakers, CER by 45--55\%, MOS remained unchanged), but did not
eliminate hallucinations on short inputs; they merely shifted to
different prompts. SFT has a structural limit here: it mimics the target
but does not actively penalize runaway behavior.

The second fine-tuning run (improved SFT) added text repetition and
short register reweighting:

\begin{itemize}
\tightlist
\item
  1 GPU, LoRA \(r=16, \alpha=32\) on all 14 layers of the backbone
  (\texttt{q,k,v,o\_proj}), LR \(2\cdot10^{-4}\), effective batch size
  64, gradient checkpointing disabled.
\item
  Everything is frozen except for the LoRA adapters (stop head, output
  heads of codec channels, audio embeddings, reference compressor, and
  unconditional prefix sampling remain frozen).
\item
  Reweight: all short lines (originally \textless{} 13 tokens) and a
  subsample of long lines yield \(\approx 27\%\) short lines in an
  epoch; selection is performed before shuffling, without duplicating
  the dataset. Long lines act as anchors against drift
  (\protect\hyperlink{ref-kumar2022lpft}{A. Kumar et al. 2022}).
\item
  LoRA was implemented natively without external adaptation libraries
  (integration, saving, and merging weights happen natively, and the
  resulting checkpoint is loaded directly into inference).
\end{itemize}

Results of the improved SFT: when using CFG, steady improvement is
observed, but clean single-pass inference (corresponding to the
production serving mode) still fails (\(\approx 27\%\) successful
generations on the most challenging slice). This motivated the
preference stage.

\hypertarget{dpo-stage}{%
\subsubsection{4.3.2. DPO Stage}\label{dpo-stage}}

SFT does not actively penalize runaway because it lacks negative
examples; DPO increases the probability of good trajectories relative to
bad ones (\protect\hyperlink{ref-rafailov2023dpo}{Rafailov et al.
2023}). The signal is accessible because the model occasionally
generates correct outputs, from which pairs can be constructed. The
pipeline consists of four steps:

\begin{verbatim}
dpo-sample -> dpo-score -> dpo-pairs -> dpo-train
\end{verbatim}

We apply length-normalized DPO (per-frame average to counteract length
bias (\protect\hyperlink{ref-meng2024simpo}{Meng, Xia, and Chen 2024})),
\(\beta = 2.0\); trajectory log-likelihood includes the stop head
Bernoulli terms (§3.3), which are mandatory for credit assignment. A key
technique is CFG distillation: two-pass text CFG is not transferred to
production (§1.2), but is used offline to generate positive examples on
difficult voices, after which its effect is baked into single-pass
weights. The complete design of reward, pairs, and DPO training, as well
as its role in eliminating short register failures, is detailed in
Section 5.

\hypertarget{training-stability-nan-incident-and-guards}{%
\subsection{4.4. Training Stability: NaN Incident and
Guards}\label{training-stability-nan-incident-and-guards}}

One of the LoRA runs progressed through 135k steps and suddenly
collapsed into NaN in a single step (from a healthy gradient norm
\(\approx 0.1\) to NaN instantly: a numerical singularity on a specific
batch rather than a learning-rate explosion). Due to checkpoint
rotation, the clean weights were lost.

Root cause:
\texttt{F.cross\_entropy(...,\ ignore\_index=-100,\ reduction=\textquotesingle{}mean\textquotesingle{})}
divides by the number of valid labels; if all labels for a head in a
microbatch are \texttt{-100}, this yields \(0/0 = \text{NaN}\),
rendering the entire loss NaN. The likely source of the degenerate audio
region was text repetition. The introduced guards include:

\begin{enumerate}
\def\labelenumi{\arabic{enumi}.}
\tightlist
\item
  \textbf{Architecture-level CE head protection:} When valid labels are
  absent in a microbatch, instead of standard cross-entropy, a weighted
  zeroing of logits is used (\texttt{logits.sum()\ *\ 0}), keeping the
  output head in the compute graph with zero gradient (similar to the
  stop loss protection, §2.5).
\item
  \textbf{Skip-on-NaN during training:} Applying the
  \texttt{torch.nan\_to\_num\_} function to gradients before the
  optimizer step.
\item
  \textbf{Filtering out degenerate samples} during data preparation.
\item
  \textbf{Backups} kept outside \texttt{output\_dir} and without the
  \texttt{checkpoint-\textless{}N\textgreater{}} prefix to prevent
  deletion by HF Trainer rotation.
\item
  \textbf{Alerts} triggered on \texttt{grad\_norm\ ==\ nan} and
  \texttt{loss\ ==\ 0.0}.
\end{enumerate}

\begin{center}\rule{0.5\linewidth}{0.5pt}\end{center}

\hypertarget{short-register-as-a-failure-mode}{%
\section{5. Short Register as a Failure
Mode}\label{short-register-as-a-failure-mode}}

The central theme of fine-tuning is the collapse of generation on short
phrases (1--2 words): runaway/never-stop. This section is structured as
failure-mode science: diagnostics rely on targeted probes rather than
the tail of the full benchmark WER. Each step is an independent
diagnostic tool designed to rule out a hypothesis, leading to the
structural root cause of the defect and proving a direct lever for
mitigation.

\hypertarget{quantitative-severity-of-the-defect}{%
\subsection{5.1. Quantitative Severity of the
Defect}\label{quantitative-severity-of-the-defect}}

We designed a targeted diagnostic stress-test for short phrases: 8 short
words (\texttt{Hello}, \texttt{Yes!}, \texttt{One.}, etc.) \(\times\) 4
voices \(\times\) 50 runs = 1600 generations, scored with
Whisper-large-v3 (\protect\hyperlink{ref-radford2022whisper}{Radford et
al. 2022}). We establish a taxonomy of failures (\texttt{timeout} /
\texttt{loop\_runaway} / \texttt{empty\_asr} / \texttt{high\_wer}),
using a \texttt{SIGALRM} timeout of 10 seconds to stop infinite loops.

On the baseline SFT model, we observed: clean 8.5\%, fail 91.5\%; 36.7\%
of generations did not return audio within 10 seconds; median duration
was 4.96 seconds for a single-syllable word; median WER was 2.0 (the
model generated \texttt{"hello\ hello\ hello..."}). The defect manifests
at the median, not in the rare tail. Across voices, the failure rate
varied from 86.8\% for the stable male voice (M1) to 98.5\% for the most
unstable female voice (F1), while the text-level variance was small
(85--97\%), indicating that the defect is systemic to the short input
class rather than tied to specific words.

\hypertarget{diagnostics-sequential-hypothesis-elimination}{%
\subsection{5.2. Diagnostics: Sequential Hypothesis
Elimination}\label{diagnostics-sequential-hypothesis-elimination}}

\hypertarget{not-a-stop-calibration-issue-p_stop-is-a-delta-function}{%
\subsubsection{\texorpdfstring{5.2.1. Not a Stop Calibration Issue:
\texttt{p\_stop} is a Delta
Function}{5.2.1. Not a Stop Calibration Issue: p\_stop is a Delta Function}}\label{not-a-stop-calibration-issue-p_stop-is-a-delta-function}}

First hypothesis: the stop head ``almost fires'' and can be fixed by
adjusting thresholds or sampling without retraining. Direct
frame-by-frame evaluation of stop probabilities \texttt{p\_stop} refutes
this. In the loop region, the distribution is binary (§3.3): median
0.0001, 99.67\% of frames are below 0.01, and exactly 0 frames fall in
the ``almost firing'' range of 0.1--0.5. Upon triggering, the value
jumps directly to 1.0.

Grouping 80 runs into modes: \texttt{clean\_stop} 21/80,
\texttt{late-stop} 28/80 (5--12 seconds of extra audio, after which the
head fires 1.0 on its own), and \texttt{never-stop} 31/80
(\texttt{p\_stop} never exceeded 0.004) (the distribution of these modes
across training rollouts is detailed in Appendix B.4). Counterfactual:
31 out of 59 runaway trajectories would not have been stopped by
stop-sampling.

The \texttt{late-stop} mode proves that the stop head is not ``blind'':
it triggers reliably on self-generated states when the backbone exits
the ``I am speaking'' mode. Consequently, the root of the runaway lies
in the backbone, not the stop head. Retraining only the
\texttt{stop\_head} is futile: a backbone LoRA is required, and the
Bernoulli terms of the trajectory (§3.3) serve as the credit assignment
channel. Diagnostics also revealed premature stops (0.2 s,
\texttt{p\_stop\ =\ 0.98}), showing that the length penalty in the
preference phase must be two-sided. Cheap inference fixes (thresholds,
stop-sampling, stop-logit temperature) are ineffective: there is nothing
to exploit between 0.004 and 1.0.

\hypertarget{not-a-bad-frame-or-phonetics-onset-probe}{%
\subsubsection{5.2.2. Not a Bad Frame or Phonetics: Onset
Probe}\label{not-a-bad-frame-or-phonetics-onset-probe}}

Second hypothesis: the collapse is caused by a single bad initial frame
or a phonetic error. We track per-frame token-NLL and belief entropy
(over 32 heads) during real inference:

\begin{longtable}[]{@{}lllll@{}}
\toprule\noalign{}
Frame & NLL Clean & NLL Derail & H Clean & H Derail \\
\midrule\noalign{}
\endhead
\bottomrule\noalign{}
\endlastfoot
0 & 41.0 & 44.8 & 1.53 & 1.62 \\
50 & 19.4 & 30.5 & 0.82 & 1.11 \\
\end{longtable}

The signature of the defect is a failure to converge. For clean
trajectories, entropy falls from 1.53 to 0.82 (the model locks onto the
speech variety); for derailed trajectories, entropy remains flat near
1.5 throughout the generation. The start is only slightly elevated
(\(\Delta \approx 3.8\) NLL at frame 0), meaning it is not a ``single
bad initial frame.'' The WER is 0 when the trajectory converges,
indicating the problem is not phonetic: the model always knows the
content. A temperature fix is ineffective (\texttt{onset\_temp\ =\ 0.2}
on the first 6 frames resulted in 51\% failures vs.~49\% without it): an
entropy of \(\approx 1.5\) represents the model's actual uncertainty at
temperature 1, and one cannot sample around genuine confusion.

\hypertarget{not-speaker-coverage-prefix-distance}{%
\subsubsection{5.2.3. Not Speaker Coverage:
Prefix-Distance}\label{not-speaker-coverage-prefix-distance}}

Third hypothesis: the defect is caused by speaker overfitting. Early DPO
rounds 1--5 on 5 voices tied the stop decision to those specific
prefixes; on a held-out external female voice (F2), the optimization
effect vanished (runaway rate was 46\%, matching the base model).
Introducing an expanded pool of 50 voices and prefix masking
(CFG-dropout at 15\%) eliminated the tie to a specific distribution: the
average held-out derail rate fell to 2.5\%. However, instability
persisted for the challenging female voice F2. Analysis of cosine
distances in the speaker space showed that the cause is not coverage:
this voice lies within the pool distribution (cosine similarity to
centroid is 0.55, compared to a pool median of 0.61), yet shows a
failure rate of 60\%. The correlation between proximity to the centroid
and failure rate is −0.05 (null). Actual geometric outliers, such as
control male voice M2 (cosine similarity 0.22), show no pronounced
instability. Consequently, simply expanding the speaker pool does not
resolve the issue for such challenging voices.

\hypertarget{structural-root-cause-and-direct-lever}{%
\subsection{5.3. Structural Root Cause and Direct
Lever}\label{structural-root-cause-and-direct-lever}}

Ruling out the three hypotheses points to the structural root cause
(§2.4): on a short input, the \(K = 8\) speaker prefix dominates over
1--2 text tokens, text conditioning is weakened, and the model fails to
lock on.

This alignment degradation highlights a fundamental difference between
decoder-only architectures (like Gepard) and encoder-decoder
architectures (like Magpie-TTS
(\protect\hyperlink{ref-nvidia2024magpie}{Neekhara et al. 2024})). In
encoder-decoder speech models, text representations are kept separate in
the encoder, and alignment is mediated by a dedicated cross-attention
mechanism. This allows researchers to apply direct loss-based
constraints (such as CTC loss or monotonic attention priors) directly
onto the cross-attention matrix to guarantee robust alignment during
training. In contrast, decoder-only models concatenate text and audio
into a single flat sequence, relying entirely on causal self-attention
to learn alignment implicitly. On short text prompts (1--2 words), this
self-attention mechanism becomes highly vulnerable: the dense query
attention from the audio frames spreads over the \(K=8\) speaker prefix
tokens, leaving almost no attention weight on the tiny text region.
Since there is no cross-attention matrix, guided attention losses cannot
be directly applied. Thus, the model's text-conditioning is
catastrophically weakened, leading to high-entropy tokens and failure to
lock on.

The direct lever to mitigate this is to boost the text weight relative
to the prefix. Text-CFG validates this lever at inference without
retraining:

\[\text{logit}_\text{guided} = \text{logit}_\text{uncond} + s\cdot(\text{logit}_\text{cond} - \text{logit}_\text{uncond})\]

\begin{longtable}[]{@{}
  >{\raggedright\arraybackslash}p{(\columnwidth - 6\tabcolsep) * \real{0.2500}}
  >{\raggedright\arraybackslash}p{(\columnwidth - 6\tabcolsep) * \real{0.2500}}
  >{\raggedright\arraybackslash}p{(\columnwidth - 6\tabcolsep) * \real{0.2500}}
  >{\raggedright\arraybackslash}p{(\columnwidth - 6\tabcolsep) * \real{0.2500}}@{}}
\toprule\noalign{}
\begin{minipage}[b]{\linewidth}\raggedright
Voice (Reference)
\end{minipage} & \begin{minipage}[b]{\linewidth}\raggedright
Without CFG (Success Rate)
\end{minipage} & \begin{minipage}[b]{\linewidth}\raggedright
With CFG (\(w = 2.6\), Success Rate)
\end{minipage} & \begin{minipage}[b]{\linewidth}\raggedright
Improvement
\end{minipage} \\
\midrule\noalign{}
\endhead
\bottomrule\noalign{}
\endlastfoot
Challenging external female voice (F2) & 7.5\% & 48.8\% & \(\times\)
6.5 \\
Target Russian speaker (RU) & 33.8\% & 61.9\% & \(\times\) 1.8 \\
Overall Average & 28.9\% & 51.6\% & \(\times\) 1.8 \\
\end{longtable}

The control female voice F2 represents a difficult case (held-out
class), resistant to text repetitions and baseline preference
optimization rounds. Increasing the success rate for this voice by 6.5
times directly validates the hypothesis that the prefix dominates over
text conditioning on ultra-short phrases. Generation durations shrink
accordingly (76 \(\to\) 36 frames): CFG pushes the model into a stable
regime. CFG is not equivalent to temperature: temperature merely
rescales a fixed distribution (which is why it is ineffective, §5.2.2),
while CFG shifts it in the direction of the text, forcing the model into
a locking regime.

\hypertarget{mitigation-two-levers}{%
\subsection{5.4. Mitigation: Two Levers}\label{mitigation-two-levers}}

The text-versus-prefix weight lever is implemented via two complementary
mechanisms: text repetition during data preparation and DPO with
CFG-generated positives at the weight level.

\hypertarget{text-repetition-budget-calibration}{%
\subsubsection{5.4.1. Text Repetition: Budget
Calibration}\label{text-repetition-budget-calibration}}

Text repetition (§4.1.5) brings the text budget of a short input up to
the stability plateau. The target budget \texttt{target\ =\ 16} was
calibrated via a stress test (simple-SVO corpus, 2 voices
\(\times \approx 80\) runs per bucket). The failure rate vs.~text budget
curve has three distinct zones: a cliff (\(\le 6\) tokens: 60--96\%
failure), a transition (7--12 tokens), and a plateau (\(\ge 13\)
tokens). The transition from 12 \(\to\) 13 tokens shows a sharp drop
(18.8\% \(\to\) 5.0\% failure); the stability plateau begins at 13
tokens for a familiar voice and at 15 for a difficult external voice,
justifying the target budget of \texttt{target\ ≈\ 16} to provide a
safety margin. At 8 tokens, a \(\approx 31\%\) failure rate remains. The
median WER is 0 in all buckets, indicating the defect is binary (failure
to converge rather than phonetics).

\begin{figure}
\centering
\includegraphics[width=1\textwidth,height=\textheight]{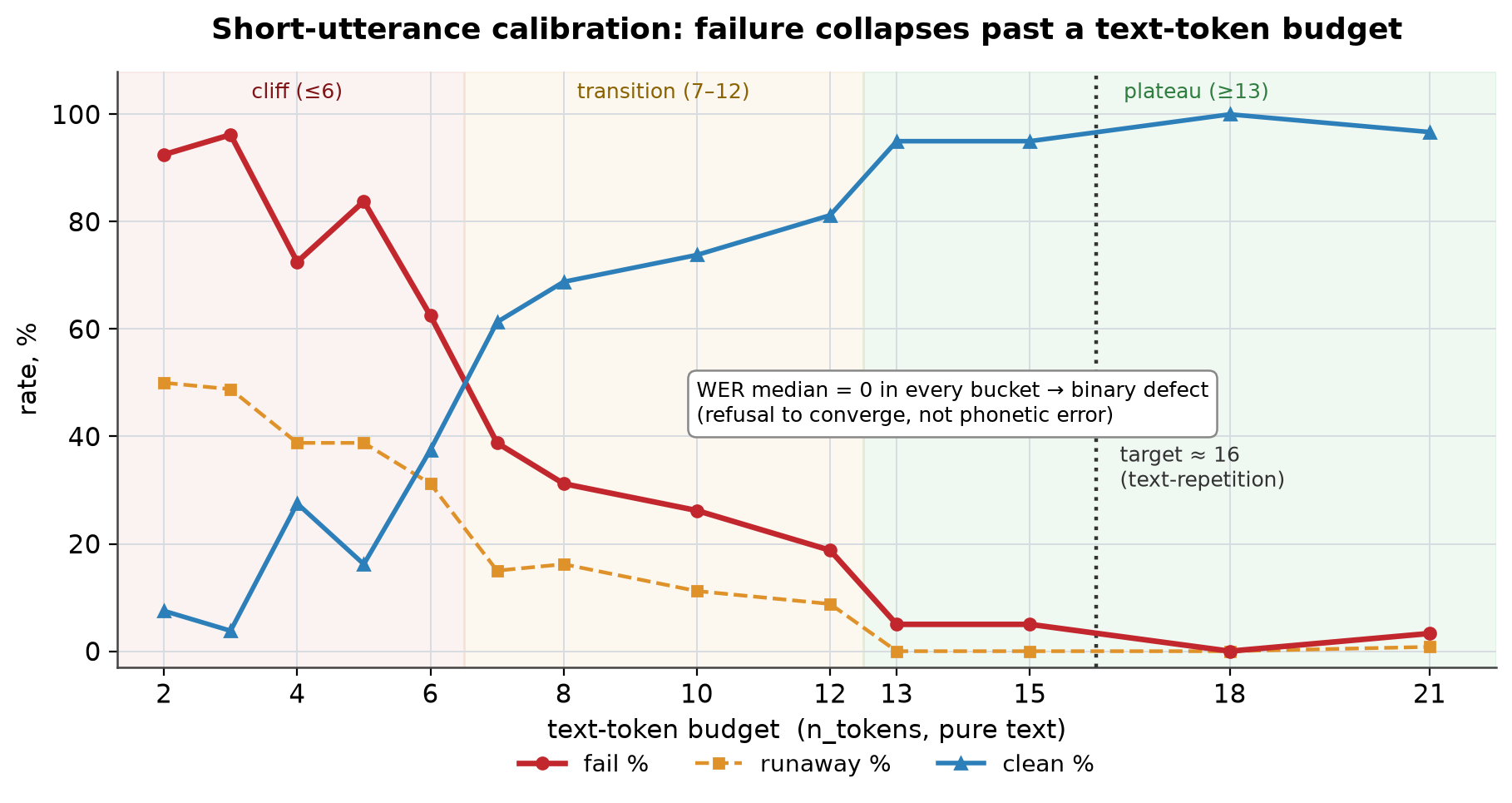}
\caption{Calibration of the target text budget for repetition
(simple-SVO corpus). The failure rate curve shows three zones (cliff /
transition / plateau); the drop at 12 \(\to\) 13 tokens and the plateau
starting at 13--15 justify the target \texttt{target\ ≈\ 16}. Median WER
is 0 in all buckets \(\to\) the defect is
binary.\label{fig:calibration}}
\end{figure}

\hypertarget{dpo-with-cfg-positives}{%
\subsubsection{5.4.2. DPO with
CFG-Positives}\label{dpo-with-cfg-positives}}

\begin{figure}
\centering
\includegraphics[width=0.62\textwidth,height=\textheight]{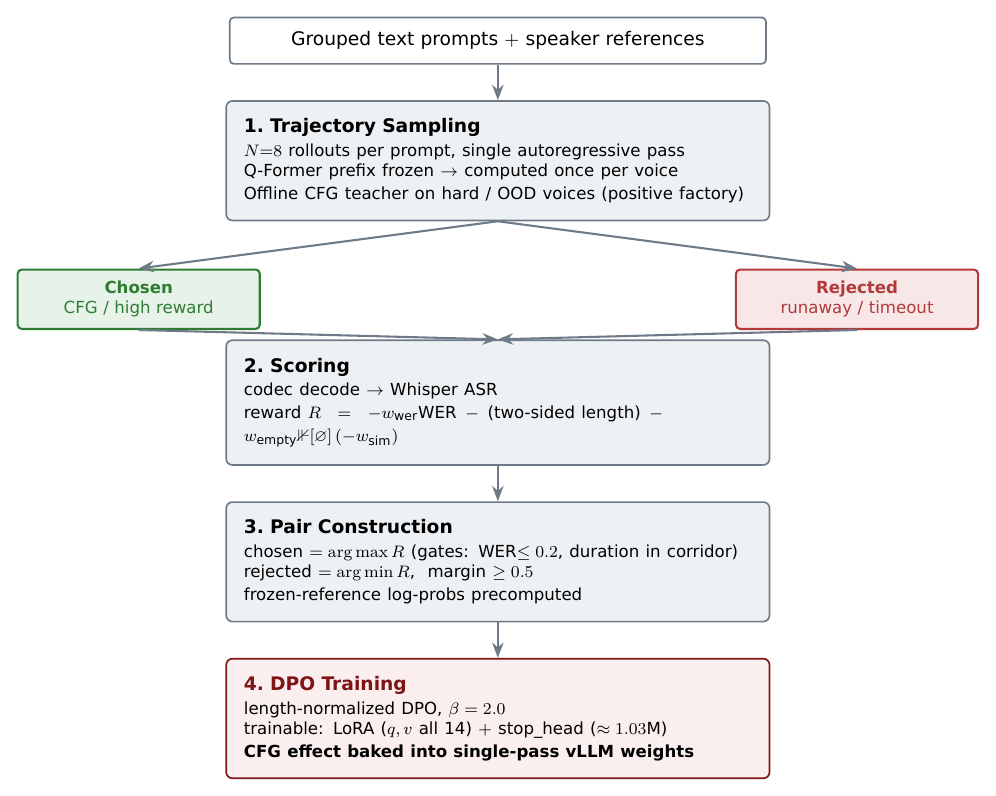}
\caption{DPO with CFG distillation. Batched rollouts (with an offline
CFG teacher on hard/OOD voices) yield chosen vs.~rejected trajectories;
an ASR-based reward with a two-sided length penalty ranks them into
pairs; length-normalized DPO bakes the two-pass CFG effect into the
single-pass weights.\label{fig:dpo}}
\end{figure}

SFT does not actively penalize runaway because it lacks negative
examples; DPO increases the probability of good trajectories relative to
bad ones. The distillation pipeline is shown in Figure \ref{fig:dpo}.

The pipeline implementation steps:

\begin{enumerate}
\def\labelenumi{\arabic{enumi}.}
\tightlist
\item
  \textbf{Trajectory Sampling:} Batch rollouts (\(N = 8\) generations
  per text prompt) are executed with an adaptive frame length limit. At
  this stage, only sampled token indices are stored (no audio files are
  saved), and voice cloning prefixes are computed once and remain
  unchanged (the Q-Former is frozen).
\item
  \textbf{Reward Evaluation:} Codec indices are reconstructed into audio
  signals and transcribed using Whisper (\texttt{distil-large-v3}). The
  trajectory reward \(R(y)\) is calculated as: \[
  \begin{aligned}
  R(y) = &- w_\text{wer} \operatorname{WER}(y) - w_\text{over} \max\bigl(0, d(y) - d_\text{max}\bigr) - w_\text{short} \mathbb{I}\bigl(d(y) < d_\text{min}\bigr) \\
         &- w_\text{empty} \mathbb{I}(\text{ASR is empty}) - w_\text{sim} \bigl(1 - \cos(y)\bigr)
  \end{aligned}
  \] The two-sided length penalty (\(w_\text{over}\) and
  \(w_\text{short}\)) is mandatory; without it, the preference
  optimization degenerates, leading the model to stop generation
  immediately at the first frame (ultra-short timing, §5.2.1).
\item
  \textbf{Pair Formulation:} The preferred candidate (chosen, \(y_w\))
  is the sample with the maximum reward \(R\) that passes logical
  filters (WER \(\le 0.2\), duration within acceptable limits, no
  truncation). The sample with the lowest reward \(R\) is assigned as
  the rejected candidate (\(y_l\)). A marginal reward threshold of
  \(\Delta R \ge 0.5\) is required. For each pair, base
  log-probabilities are calculated on the frozen reference model.
\item
  \textbf{Training:} Optimization is conducted using the
  length-normalized DPO criterion (§A.6) at \(\beta = 2.0\)
  (\protect\hyperlink{ref-meng2024simpo}{Meng, Xia, and Chen 2024}). The
  trajectory log-likelihood is the sum of the log-probabilities from the
  32 heads and the stop head's Bernoulli terms. Due to stop-probability
  saturation, the logical \(p_{\text{stop}}\) value is clipped from
  below at \(p_{\text{floor}} = 10^{-4}\) to ensure gradient signal
  propagation (§3.3). Approximately \(1.03\text{M}\) parameters are
  optimized (LoRA adapters on the \texttt{q,v} projections of all 14
  layers of the transformer and the weights of the stop head). The
  output projections of the codec channels and the base weights remain
  frozen.
\end{enumerate}

\emph{CFG Distillation.} At inference, CFG is not deployed in production
(two passes conflict with vLLM's continuous batching, §1.2), but it is
used offline to generate positive examples for difficult voices (with a
scale of \(w_\text{cfg} = 3\), onset-only 20 frames), where standard
sampling yields only negative trajectories. The CFG effect is baked into
single-pass weights via DPO, leaving inference single-pass.

\hypertarget{results}{%
\subsection{5.5. Results}\label{results}}

The stress test is deliberately configured with a worst-case scenario:
single-word inputs \(\times\) 5 external difficult voices, single-pass
inference (no CFG, production-equivalent for vLLM), temperature 0.4.

\begin{longtable}[]{@{}
  >{\raggedright\arraybackslash}p{(\columnwidth - 6\tabcolsep) * \real{0.2500}}
  >{\raggedright\arraybackslash}p{(\columnwidth - 6\tabcolsep) * \real{0.2500}}
  >{\raggedright\arraybackslash}p{(\columnwidth - 6\tabcolsep) * \real{0.2500}}
  >{\raggedright\arraybackslash}p{(\columnwidth - 6\tabcolsep) * \real{0.2500}}@{}}
\toprule\noalign{}
\begin{minipage}[b]{\linewidth}\raggedright
Metric
\end{minipage} & \begin{minipage}[b]{\linewidth}\raggedright
Start (Baseline SFT)
\end{minipage} & \begin{minipage}[b]{\linewidth}\raggedright
Improved SFT
\end{minipage} & \begin{minipage}[b]{\linewidth}\raggedright
Final (DPO, Round 3)
\end{minipage} \\
\midrule\noalign{}
\endhead
\bottomrule\noalign{}
\endlastfoot
Clean & 26.8\% & 26.8\% & 67.1\% (\(\approx 71\%\) @ t0.3) \\
Runaway + Timeout & \(\approx\) 55\% & 25.4\% timeout & \(\approx\) 5\%
(timeout 0) \\
WER mean & 5.76 & 5.76 & 1.56 \\
Female Voice F1 (Challenging) & 0.6\% & 0.6\% & 55.0\% \\
\end{longtable}

The single-pass clean generation rate grows across rounds: 26.8\%
\(\to\) 52.9\% (Round 1) \(\to\) 63.5\% (Round 2) \(\to\) 67.1\% (Round
3), exceeding the CFG ceiling of the improved SFT (64.8\%), which
confirms the success of distilling the CFG effect into the model
weights. On the full diagnostic set after 3 rounds of DPO, the failure
rate dropped by \(\approx 25\) times (for the control male voice M2:
9.8\% \(\to\) 0.4\%, for the target Russian speaker RU: 12.2\% \(\to\)
0.9\%), MOS increased by \(+0.22\dots+0.29\), and the runaway effect was
almost eliminated. For long texts, stability is fully resolved (0\%
derail rate across all voices, WER 0.14--0.18, MOS up to 4.94), making
this mode production-ready. The remaining defect, elevated WER on short
inputs, reflects sampling variance and instability rather than a gap in
the model's knowledge.

The effectiveness of each successive optimization step decreases
monotonically (+26.1 \(\to\) +10.6 \(\to\) +3.6 percentage points across
rounds), pairs become sparser, and the offline teacher with CFG is near
exhaustion. We select the second-round model (DPO-r2) as the final
configuration: the third round (DPO-r3) causes undesirable drift on OOD
prompts (as discussed below), while DPO-r2 maintains an optimal balance.
This table reflects the results of the internal stress diagnostics
(monosyllabic inputs \(\times\) difficult voices) and is reported
separately from the public benchmark results (Section 6).

\emph{Regressions and Lessons:}

\begin{itemize}
\tightlist
\item
  DPO drifts on OOD prompts (categories outside the training set:
  casual, URLs, emotional) because the KL anchor acts only on the
  distribution of pairs. This is mitigated by data coverage (anchor
  corpus v2) rather than reward modification.
\item
  Speaker-overfit stop decisions (Rounds 1--5 on 5 voices) tied the stop
  to those 5 prefixes. This is resolved using a diverse pool (50 voices)
  and a \texttt{null\_prefix} CFG (Round 6: held-out derail rate 49\%
  \(\to\) 2.5\%, §5.2.3).
\end{itemize}

\begin{center}\rule{0.5\linewidth}{0.5pt}\end{center}

\hypertarget{inference-and-evaluation}{%
\section{6. Inference and Evaluation}\label{inference-and-evaluation}}

Deployment is the primary driver of the architecture (§1). This section
describes inference runtimes (§6.1), speed measurements (§6.2), and
quality on the public Seed-TTS-eval (§6.3). Speed and quality are
measured on different sets; each number is explicitly marked with its
corresponding dataset.

\hypertarget{deployment-runtimes}{%
\subsection{6.1. Deployment Runtimes}\label{deployment-runtimes}}

Production serving uses vLLM (\protect\hyperlink{ref-kwon2023vllm}{Kwon
et al. 2023}). The backbone is a standard full-attention transformer and
is served by vLLM without custom patches to the engine. The mode is
single-pass (one AR pass per frame): all non-standard processing is
moved out of the decode loop. The Q-Former prefix is computed once
during prefill (the frozen compressor outputs a constant speaker
representation) and is passed as \texttt{prompt\_token\_ids} or raw
embeddings; text repetition is implemented only during prompt/prefill
and does not touch the engine.

The reference runtime based on standard libraries is used for
diagnostics and offline tasks in two modes. Single-pass inference (no
guidance) yields the same distribution as the serving environment and is
used for honest benchmarks and generating DPO trajectories; an optional
length-limiting mechanism can force stops by frame counts. Two-pass
inference with CFG performs classifier-free guidance over the text
relative to the prefix with two passes per frame:

\[\text{cond} = [\text{prefix} \mid \text{text} \mid \text{audio}], \qquad \text{uncond} = [\text{prefix} \mid \varnothing\text{text} \mid \text{audio}]\]
\[\text{logit}_\text{guided} = \text{logit}_\text{uncond} + s\cdot(\text{logit}_\text{cond} - \text{logit}_\text{uncond})\]

with \(s = 3\) and onset-only guidance of \(\approx 20\) frames. The CFG
mode is not transferred to production, as two passes and sequence
binding conflict with continuous batching, and is used solely offline to
generate successful trajectories for DPO training (§5.4.2).

\hypertarget{speed}{%
\subsection{6.2. Speed}\label{speed}}

Speed was measured in two stages with different objectives (§1.3). Stage
1 (sanity) was a single-stream vLLM run on a rented RTX 5090 (Vast.ai),
without environment controls, yielding an RTF of \(\approx 0.040\) and a
TTFA of \(\approx 0.032\text{ s}\) (about 25\(\times\) faster than
real-time). This stage is not a controlled benchmark but confirmed the
order of magnitude and removed the risk of the vLLM approach failing to
meet real-time constraints.

Stage 2 (realistic serving) is an end-to-end benchmark (backbone and
neural codec) via SSE streaming under concurrent load on a server-class
GPU (specifically, an AWS g7e.2xlarge instance equipped with a 96 GB
NVIDIA RTX PRO 6000 Blackwell GPU): full CUDA graph and embed CUDA
graph, \texttt{max\_num\_seqs\ =\ 256},
\texttt{gpu\_memory\_utilization\ =\ 0.82},
\texttt{CODEC\_MAX\_DECODE\_BATCH\ =\ 32}, bf16, temperature 0.3, and
\texttt{max\_tokens\ =\ 400}. Measurement is conducted along the
concurrency axis \(C = 1\dots256\); TTFB is the wall time to the first
audio delta, and RTF is the per-stream ratio of request wall time to
audio duration (\(\text{wall}_i / \text{audio}_i\), where \(< 1\) is
faster than real-time). The aggregate system speedup is computed as
\(\text{xRT} = \sum_i \text{audio}_i / \text{wall}_{\text{elapsed}}\),
where \(\text{wall}_{\text{elapsed}}\) is the total elapsed time for the
concurrent batch run. To eliminate noise from stochastic early stops
(where a request terminates prematurely and the fixed TTFB overhead
skews the RTF), streams yielding less than \(2.0\text{ s}\) of audio are
filtered out from the statistics. Detailed benchmarking conditions,
equations, and filtering thresholds are described in Appendix A.7.

\begin{longtable}[]{@{}
  >{\raggedright\arraybackslash}p{(\columnwidth - 14\tabcolsep) * \real{0.1250}}
  >{\raggedright\arraybackslash}p{(\columnwidth - 14\tabcolsep) * \real{0.1250}}
  >{\raggedright\arraybackslash}p{(\columnwidth - 14\tabcolsep) * \real{0.1250}}
  >{\raggedright\arraybackslash}p{(\columnwidth - 14\tabcolsep) * \real{0.1250}}
  >{\raggedright\arraybackslash}p{(\columnwidth - 14\tabcolsep) * \real{0.1250}}
  >{\raggedright\arraybackslash}p{(\columnwidth - 14\tabcolsep) * \real{0.1250}}
  >{\raggedright\arraybackslash}p{(\columnwidth - 14\tabcolsep) * \real{0.1250}}
  >{\raggedright\arraybackslash}p{(\columnwidth - 14\tabcolsep) * \real{0.1250}}@{}}
\toprule\noalign{}
\begin{minipage}[b]{\linewidth}\raggedright
\(C\) (Streams)
\end{minipage} & \begin{minipage}[b]{\linewidth}\raggedright
TTFB p50
\end{minipage} & \begin{minipage}[b]{\linewidth}\raggedright
TTFB p90
\end{minipage} & \begin{minipage}[b]{\linewidth}\raggedright
TTFB p99
\end{minipage} & \begin{minipage}[b]{\linewidth}\raggedright
RTF p50
\end{minipage} & \begin{minipage}[b]{\linewidth}\raggedright
RTF p90
\end{minipage} & \begin{minipage}[b]{\linewidth}\raggedright
RTF p99
\end{minipage} & \begin{minipage}[b]{\linewidth}\raggedright
xRT (System)
\end{minipage} \\
\midrule\noalign{}
\endhead
\bottomrule\noalign{}
\endlastfoot
1 & 0.046 s & 0.046 s & 0.046 s & 0.067 & 0.067 & 0.067 &
15.0\(\times\) \\
32 & 0.157 s & 0.280 s & 0.304 s & 0.221 & 0.273 & 0.283 &
116.1\(\times\) \\
128 & 0.342 s & 0.607 s & 0.697 s & 0.710 & 0.803 & 0.853 &
168.3\(\times\) \\
256 & 0.716 s & 0.956 s & 0.990 s & 1.214 & 1.372 & 1.503 &
203.9\(\times\) \\
\end{longtable}

The table corresponds to the optimal inference configuration (with
generation block size \texttt{CHUNK=86}, polling interval
\texttt{FETCH=86}, and initial buffer \texttt{FIRST=10}), balancing
latency (TTFB) and throughput. Compared to baseline settings, the
optimized parameters provide a twofold reduction in latency to the first
frame on a single stream, and the peak aggregate speedup reaches
203.9\(\times\). A full sweep of four streaming configurations (default,
low-latency, throughput-oriented, and balanced) is provided in
\texttt{speed\_test/}; a throughput-oriented configuration with larger
chunks raises the \(C=256\) aggregate to \(\approx 224\times\), at the
cost of higher TTFB.

Conclusions: - A single stream runs with a real-time safety margin: RTF
\(\approx 0.067\), about 15\(\times\) faster than real-time. -
Throughput scales up to \(C = 256\) (peak xRT \(\approx 204\times\) in
this configuration) without saturation, indicating the system is
compute-bound with headroom rather than hitting a memory wall. -
Per-stream RTF crosses 1.0 between \(C = 128\) and \(C = 256\): at
\(C \le 128\), each stream is faster than real-time (RTF
\(\le \approx 0.7\)), whereas at \(C = 256\), the GPU is overloaded and
a single stream slows to \(\approx 1.1\text{--}1.2\). The practical
operating range for interactive serving is \(C \approx 64\text{--}128\);
\(C = 256\) is justified only for latency-tolerant batch processing. -
The pipeline is LLM-bound rather than codec-bound: increasing the
codec's decoding batch size does not yield performance gains (a negative
result), and synchronous decoding merely delays backbone execution.

\hypertarget{quality-seed-tts-eval}{%
\subsection{6.3. Quality: Seed-TTS-eval}\label{quality-seed-tts-eval}}

Quality was measured on the public Seed-TTS-eval
(\protect\hyperlink{ref-bytedance2024seedtts}{Anastassiou et al. 2024}),
using 1088 paired prompts (identical UUIDs and texts across all models).
The main comparison here is against the \textbf{open-source
voice-cloning systems} we positioned against from the start (§1.4): six
baselines run with reference cloning (VoxCPM2, Fish-S2, OmniVoice,
Qwen3-TTS, Echo-TTS, Chatterbox) plus Gepard. All seven were synthesized
with reference audio (\texttt{seed\_tts\_eval\_en\_vc} mode), so
\textbf{speaker similarity (SIM) is defined for the whole cohort} and
the comparison is fully paired --- our own runs on identical inputs, not
a compilation of third-party numbers. A broader comparison against
large-scale and commercial systems (Cartesia, ElevenLabs, Grok-TTS,
Kokoro, VibeVoice), which were run without cloning and therefore have no
SIM, is reported separately in \textbf{Appendix C}.

Gepard was deliberately run in single-pass mode without CFG, the
production-equivalent path (what is actually deployed on vLLM, §6.1)
rather than a CFG-enhanced diagnostic mode. The comparison is thus
conservative for Gepard: CFG mode would yield better numbers, but it is
not the release candidate. Metrics: WER/CER (Whisper
(\protect\hyperlink{ref-radford2022whisper}{Radford et al. 2022}),
intelligibility), SIM (WavLM (\protect\hyperlink{ref-chen2021wavlm}{Chen
et al. 2021}), voice similarity), UTMOS
(\protect\hyperlink{ref-saeki2022utmos}{Saeki et al. 2022})
(naturalness), and NISQA-MOS
(\protect\hyperlink{ref-mittag2021nisqa}{Mittag et al. 2021}) (overall
signal quality), together with three interpretable NISQA signal
dimensions reported in the tables: \textbf{NOI} (noisiness --- higher
means less background noise), \textbf{COL} (coloration --- higher means
less timbral/spectral distortion), and \textbf{DIS} (discontinuity ---
higher means fewer glitches and dropouts). All NISQA scores are on a
1--5 scale where higher is better.

\begin{table}[h]
\centering
\footnotesize
\setlength{\tabcolsep}{5pt}
\renewcommand{\arraystretch}{1.15}
\begin{tabular}{lccccccc}
\hline
\textbf{Model} & WER\,$\downarrow$ & SIM\,$\uparrow$ & UTMOS\,$\uparrow$ & NISQA-MOS\,$\uparrow$ & NOI\,$\uparrow$ & COL\,$\uparrow$ & DIS\,$\uparrow$ \\
\hline
VoxCPM2 & \textbf{0.015} & \textbf{0.867} & 2.42 & 3.97 & 3.86 & 3.96 & 4.30 \\
Fish-S2 & 0.016 & 0.789 & 2.80 & 4.18 & 3.87 & 4.14 & 4.44 \\
OmniVoice & 0.016 & 0.848 & 2.63 & 4.17 & 4.14 & 4.13 & 4.44 \\
Qwen3-TTS & 0.017 & 0.833 & \textbf{2.87} & 4.18 & 3.89 & 4.14 & 4.43 \\
Echo-TTS & 0.022 & 0.824 & 2.60 & 4.08 & 3.78 & 4.07 & 4.36 \\
\hdashline
\textbf{Gepard (Ours)} & 0.036 & 0.585 & 2.64 & \textbf{4.25} & \textbf{4.16} & \textbf{4.16} & \textbf{4.51} \\
\hdashline
Chatterbox & 0.063 & 0.796 & 2.70 & 4.19 & 4.12 & 4.12 & 4.46 \\
\hline
\end{tabular}
\end{table}

Sorted by WER. Best value in each column is bold; the broader field of
commercial / large-scale systems is in Appendix C.

Interpretation: - \textbf{NISQA-MOS 4.25 --- the best in the cohort}
(next: Chatterbox 4.19, Fish-S2 / Qwen3-TTS 4.18), and Gepard also leads
every interpretable NISQA quality dimension except loudness: \textbf{NOI
4.16 (noisiness), COL 4.16 (coloration), and DIS 4.51 (discontinuity)
are all the highest}. Among the open-source voice-cloning systems,
Gepard produces the cleanest, least-noisy, least-distorted, and
smoothest signal: on perceptual signal quality the model leads its
field. - \textbf{WER 0.036} is in the lower-middle of the cohort (6th of
7): better than Chatterbox (0.063), worse than the dense top group
(\(\approx 0.015\text{--}0.022\)). The result is below SOTA
intelligibility but consistent with our positioning (§1.4): this work
does not target SOTA WER. - \textbf{SIM 0.585} is the lowest in the
cohort (others 0.79--0.87). It directly confirms the voice-cloning
defect (train-time representation leakage, §4.2) and is the main
weakness of the current version: the issue is speaker similarity, not
signal quality.

\begin{figure}
\centering
\includegraphics[width=1\textwidth,height=\textheight]{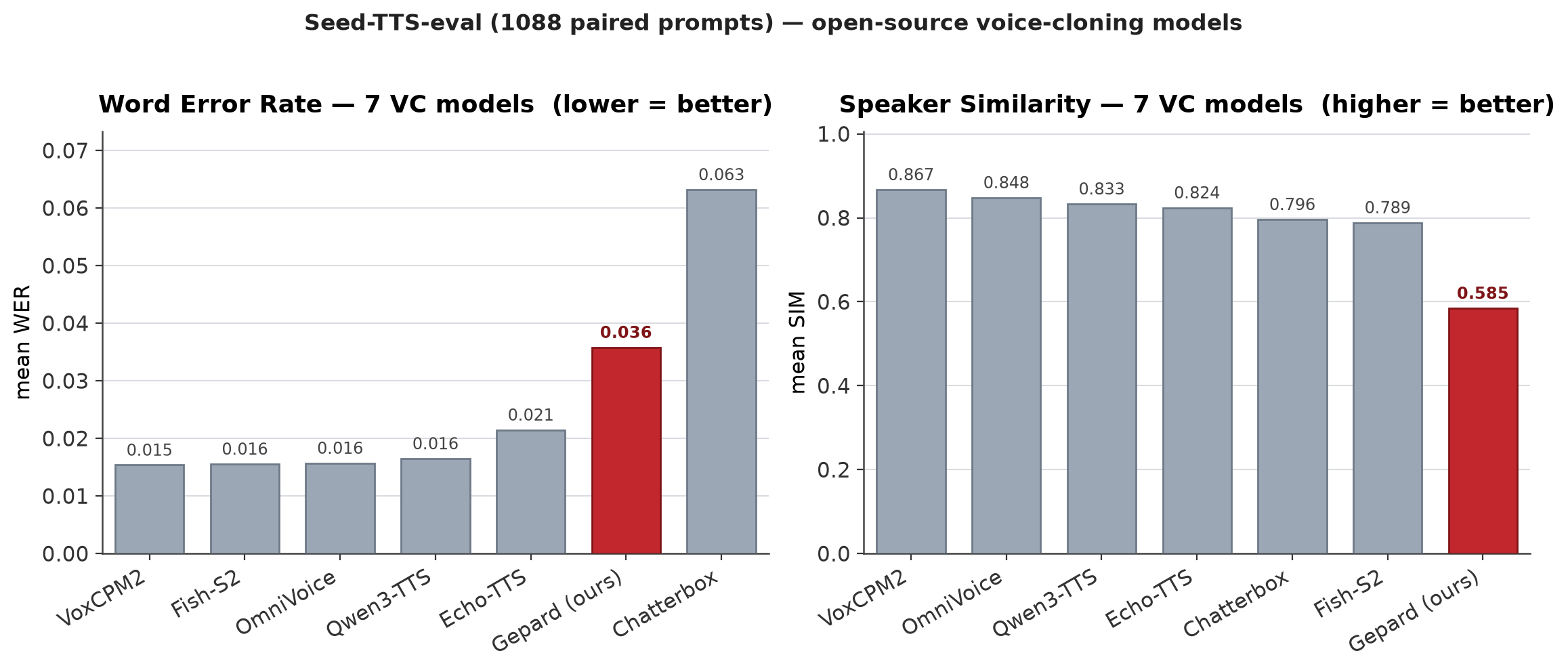}
\caption{Seed-TTS-eval (1088 paired prompts), open-source voice-cloning
cohort (7 models): intelligibility (WER\(\downarrow\)) and voice
similarity (SIM\(\uparrow\)). Gepard shows acceptable WER but the lowest
SIM (main weakness, representation leakage).\label{fig:bench-wer-sim}}
\end{figure}

\begin{figure}
\centering
\includegraphics[width=1\textwidth,height=\textheight]{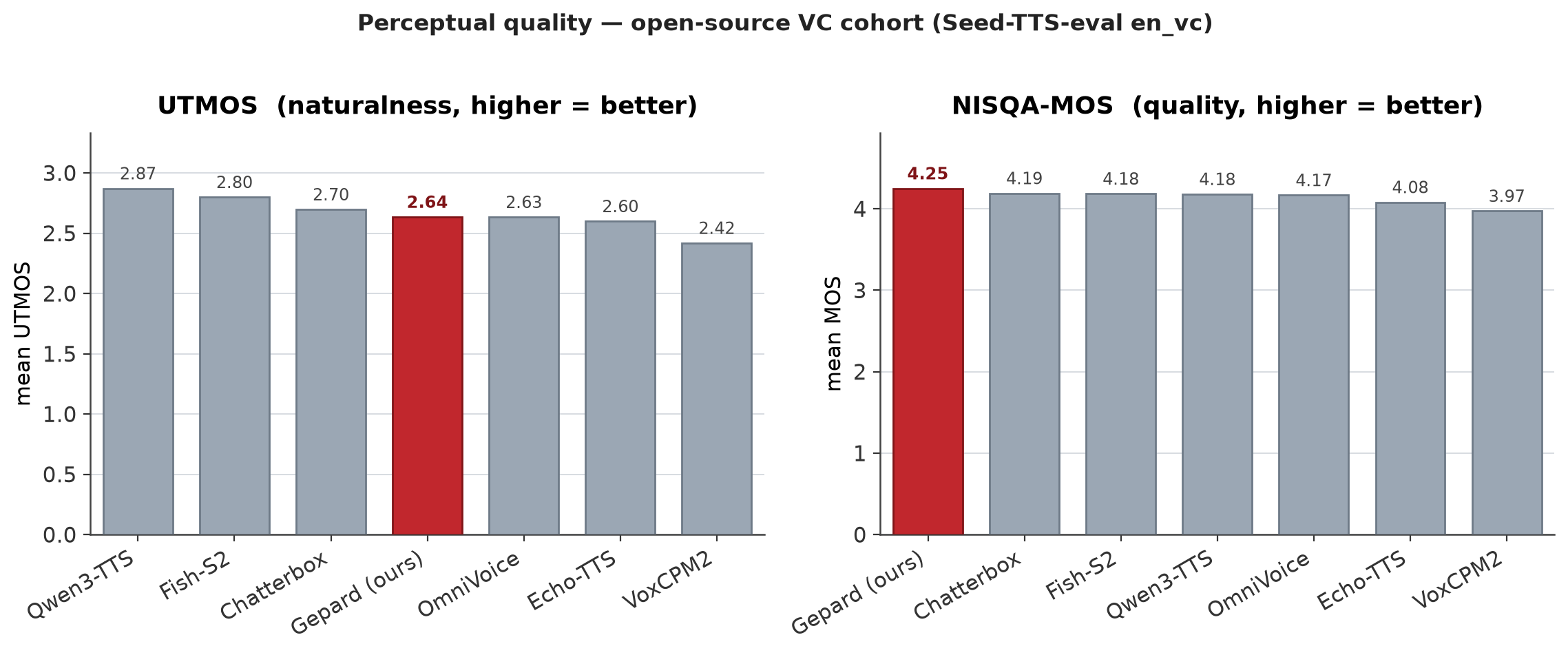}
\caption{Seed-TTS-eval: naturalness (UTMOS\(\uparrow\)) and signal
quality (NISQA-MOS\(\uparrow\)) across the open-source VC cohort. Gepard
has the highest NISQA-MOS, showing that the issue lies in speaker
similarity rather than signal quality.\label{fig:bench-utmos-mos}}
\end{figure}

\begin{figure}
\centering
\includegraphics[width=0.8\textwidth,height=\textheight]{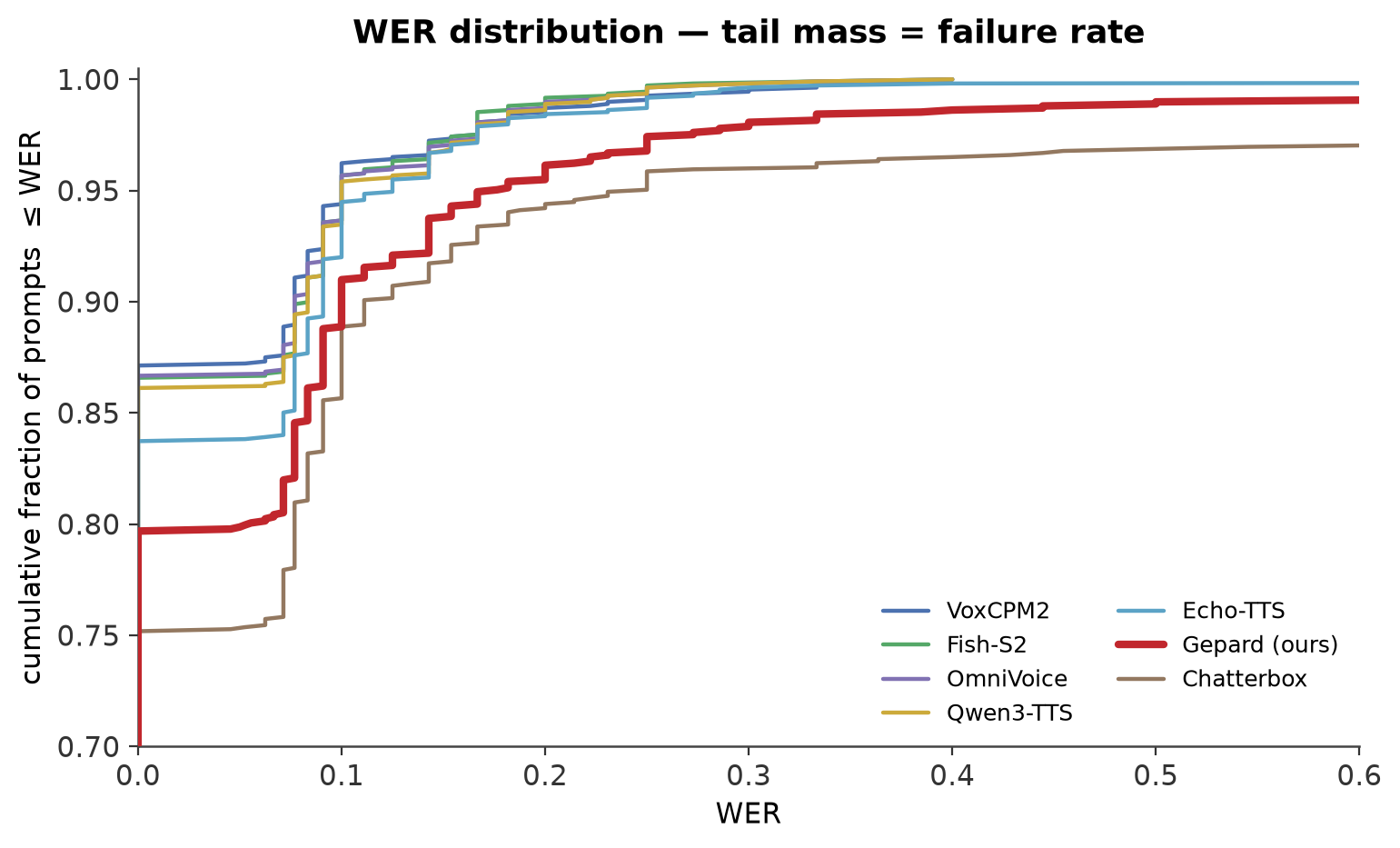}
\caption{CDF of WER across the open-source VC cohort; the right tail
corresponds to the failure rate (WER \textgreater{} 0.5). Gepard's tail
is thinner than Chatterbox but thicker than the top
group.\label{fig:bench-wer-cdf}}
\end{figure}

Length coverage limits interpretation. Seed-TTS-eval contains almost no
1--2 word prompts (minimum 3 words, median 13 tokens), so the primary
failure mode of this project (Section 5) is under-tested by this
benchmark. This explains why the WER here is 0.036, whereas the
adversarial slice in §5.5 showed a complete collapse: the numbers refer
to different evaluation sets, as labeled.

\begin{figure}
\centering
\includegraphics[width=0.8\textwidth,height=\textheight]{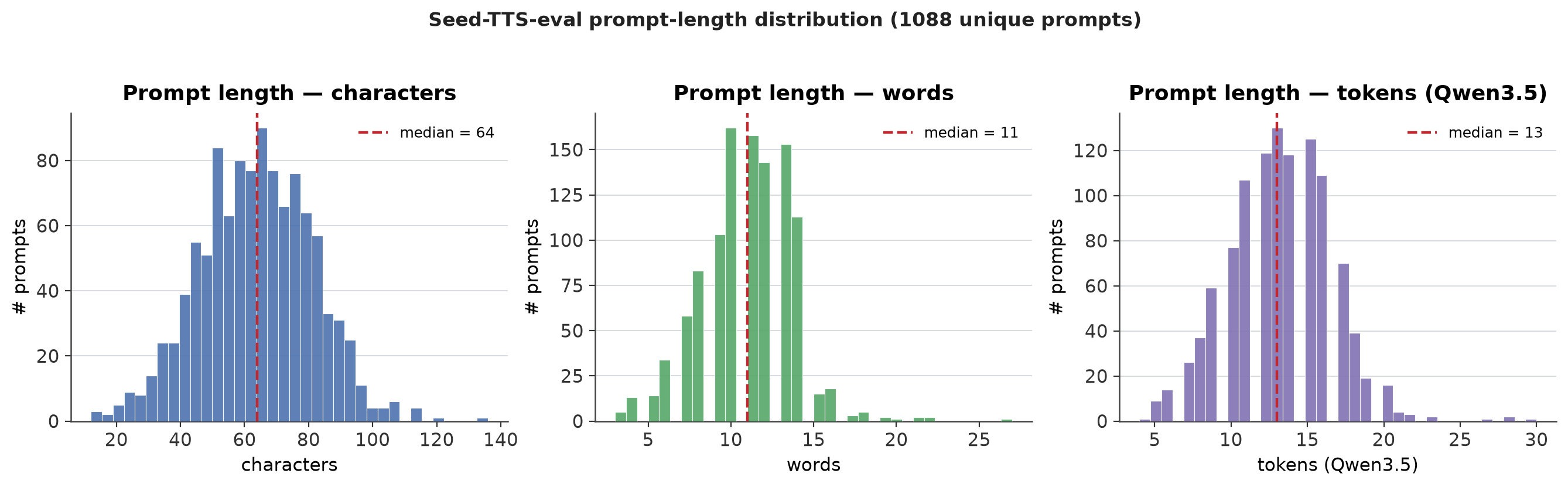}
\caption{Prompt length distribution in Seed-TTS-eval (characters / words
/ tokens). Most mass lies above the 1--2 word regime (median 13 tokens),
so the short register is barely covered by this
dataset.\label{fig:text-length}}
\end{figure}

We note a qualitative observation (not formally measured): the model
sounds natural, with realistic intonation, breathing, and rhythm.
Metrics like UTMOS/NISQA/WER do not capture this ``liveliness''
directly; NISQA-MOS 4.25 (best in the VC cohort) serves as a partial
proxy. This qualitative claim is a candidate for human-MOS/preference
evaluation on a subset and is presented as a qualitative result.

\begin{center}\rule{0.5\linewidth}{0.5pt}\end{center}

\hypertarget{limitations-and-future-work}{%
\section{7. Limitations and Future
Work}\label{limitations-and-future-work}}

\hypertarget{limitations-of-the-study}{%
\subsection{7.1. Limitations of the
Study}\label{limitations-of-the-study}}

Despite the achieved results in inference speed and stability, the
current version of Gepard (v0.0.1) has several limitations:

\begin{enumerate}
\def\labelenumi{\arabic{enumi}.}
\tightlist
\item
  \textbf{Representation Leakage in the Voice Compressor (Speaker
  Similarity vs.~Leakage):} The low WavLM-similarity score
  (\(\text{SIM} = 0.585\)) is the most critical limitation. The GroupFSQ
  quantization mechanism combined with the Q-Former has a strong
  gradient incentive to copy spectral noise and acoustic characteristics
  of the reference clip. SupCon regularization partially addresses this
  during training but does not guarantee full timbre transfer to unseen
  voices, keeping the current Voice Cloning mode in a Proof of Concept
  status.
\item
  \textbf{Lack of Strict Controlled Ablation Studies:}

  \begin{itemize}
  \tightlist
  \item
    \textbf{Backbone Architecture:} The decision to remove
    \texttt{LinearBlock} layers from Qwen3.5 and retain only classic
    full attention was based on qualitative evaluation (subjective
    listening) and not backed by quantitative metrics. The impact of
    this modification on the representation capacity of the backbone
    remains unstudied.
  \item
    \textbf{Quantization Depth (Depth Head):} The theoretical argument
    regarding the redundancy of a depth-transformer over GroupFSQ codes
    (§3.2) lacks experimental validation in the form of ablation tests
    (comparing training with and without a depth head).
  \end{itemize}
\item
  \textbf{Residual Instability on Ultra-Short Phrases:} Stabilizing the
  short register using text repetitions and DPO distillation is an
  heuristic solution. For the challenging female voice F1, the failure
  rate remains high (55\%), indicating pathological incompatibility of
  certain reference embeddings with text prefixes. The decoder-only TTS
  concept with a stop head likely has fundamental length limitations
  that require explicit length-conditioning or positional encodings such
  as PM-RoPE (\protect\hyperlink{ref-voicestar2025}{Peng et al. 2025}).
\item
  \textbf{Language and Domain Imbalance:} The limited volume of training
  data in non-English languages led to a pronounced degradation of the
  audio interface outside the English domain, restricting the practical
  application of the current version for multilingual synthesis.
\end{enumerate}

\hypertarget{future-research-directions}{%
\subsection{7.2. Future Research
Directions}\label{future-research-directions}}

To address the identified limitations and develop the Gepard
architecture, the following steps are planned:

\begin{enumerate}
\def\labelenumi{\arabic{enumi}.}
\tightlist
\item
  \textbf{Scaling Voice Cloning:} Transitioning to full compressor
  training on unrelated pairs (different clips of the same speaker as
  reference and target) to eliminate representation leakage, and
  expanding the pool of evaluated unseen speakers.
\item
  \textbf{Quantitative Architecture Validation:} Conducting systematic
  ablation experiments comparing the hybrid backbone (retaining
  \texttt{LinearBlock}) with full-attention, and quantitatively
  evaluating the role of the depth-head projector in decoding codec
  codes.
\item
  \textbf{Refining Stop Mechanisms:} Investigating alternative length
  conditioning methods (e.g., explicit duration prediction from text
  during prefill, or attention modifications via Attention Guidance
  (\protect\hyperlink{ref-attentionguidance2025}{S. Wang et al. 2025})).
\item
  \textbf{Preference Optimization via GRPO:} Moving from pairwise DPO to
  Group Relative Policy Optimization (GRPO)
  (\protect\hyperlink{ref-shao2024deepseekmath}{Shao et al. 2024}). GRPO
  is an on-policy reinforcement learning algorithm that optimizes policy
  behavior by comparing rewards within a generated group of outputs for
  the same prompt, rather than relying on a separate reference model for
  KL-regularization. This is highly aligned with our multi-sample
  rollout pipeline (\(N=8\)), dramatically reduces GPU memory
  requirements by eliminating the need to load the reference model
  during training, and allows direct optimization of speech quality and
  stability metrics (like WER and speaker similarity) via relative
  reward signals.
\item
  \textbf{Multilingual Alignment:} Expanding the non-English portion of
  the training corpus and conducting targeted audio interface
  fine-tuning to achieve quality parity across all languages supported
  by the backbone.
\end{enumerate}

\begin{center}\rule{0.5\linewidth}{0.5pt}\end{center}

\hypertarget{release-and-reproducibility}{%
\section{8. Release and
Reproducibility}\label{release-and-reproducibility}}

The primary release artifact is the trained parameter chain:

\begin{verbatim}
Base pretrain model (555.7M parameters)
        +- + LoRA-SFT (text repetition + length reweighting, §4.3.1)
                +- + DPO (final model after 2nd round of preference optimization)
\end{verbatim}

The second-round DPO model (DPO-r2) is selected as the final
configuration: the third round leads to undesirable drift on OOD prompts
(§5.5), while the second round maintains an optimal balance of stability
and quality. The neural codec is
\texttt{nvidia/nemo-nano-codec-22khz-1.89kbps-21.5fps} (§2.3). Inference
behavior is safeguarded by an optional deterministic frame limit. Model
parameters, data license specifications, a demo page with audio samples,
and the optimized serving solution are published in the open domain.

\hypertarget{summary-of-hyperparameters-and-configuration-heuristics}{%
\subsection{8.1. Summary of Hyperparameters and Configuration
Heuristics}\label{summary-of-hyperparameters-and-configuration-heuristics}}

\begin{longtable}[]{@{}
  >{\raggedright\arraybackslash}p{(\columnwidth - 4\tabcolsep) * \real{0.3333}}
  >{\raggedright\arraybackslash}p{(\columnwidth - 4\tabcolsep) * \real{0.3333}}
  >{\raggedright\arraybackslash}p{(\columnwidth - 4\tabcolsep) * \real{0.3333}}@{}}
\toprule\noalign{}
\begin{minipage}[b]{\linewidth}\raggedright
Component / Parameter Set
\end{minipage} & \begin{minipage}[b]{\linewidth}\raggedright
Scope
\end{minipage} & \begin{minipage}[b]{\linewidth}\raggedright
Key Hyperparameters and Heuristics
\end{minipage} \\
\midrule\noalign{}
\endhead
\bottomrule\noalign{}
\endlastfoot
Backbone and Audio Output Architecture & Heads and Stop Predictor
Configuration & 32 audio heads (\([8,7,6,6]\times 8\)); stop loss weight
\(w_{\text{stop}}=2.0\), positive class weight
\(w_{\text{pos}}=25.0\) \\
Data Representation and Augmentations & Codec, Filtering, and Repetition
Policies & NanoCodec (\(8 \times [8,7,6,6]\) channels, \(21.5\)
frames/s); active \texttt{text\_repetition} (target budget of 16
tokens) \\
Speaker Conditioning (Voice Cloning) & Q-Former, CFG-Dropout, and SupCon
& \(K=8\) latent queries, \(L=2\) layers; CFG-dropout \(0.15\);
compressor parameters frozen during fine-tuning \\
Pretraining & Base Model Optimization & 8 epochs, base learning rate
(LR) \(9\cdot 10^{-4}\) (cosine decay); LR scaling for audio codec
\(\times 0.5\) / text embedding \(\times 0.2\); FSDP2 \\
Fine-Tuning (LoRA-SFT) & Short Register Adaptation & Frozen base
backbone + LoRA (\(r=16, \alpha=32\)) on all 14 layers of the
transformer (\texttt{q,k,v,o}); LR \(2\cdot 10^{-4}\), effective batch
size 64 \\
Preference Optimization (DPO) & Distillation and Stabilization & DPO
coefficient \(\beta=2.0\) with length normalization; LoRA adapters on
projections (\texttt{q,v}) of all 14 layers + stop head training;
two-sided length reward \\
\end{longtable}

\begin{center}\rule{0.5\linewidth}{0.5pt}\end{center}

\hypertarget{appendix-a.-mathematical-description-and-implementation-details}{%
\section{Appendix A. Mathematical Description and Implementation
Details}\label{appendix-a.-mathematical-description-and-implementation-details}}

The main text provides the key concepts; they are systematized here by
component to improve reproducibility.

\hypertarget{a.1.-codec-unfold-and-dequantize}{%
\subsection{A.1. Codec: Unfold and
Dequantize}\label{a.1.-codec-unfold-and-dequantize}}

In the codec decoder implementation (§2.3), the packed token of the
codebook is unfolded into \(D = 4\) independent FSQ channels via
mixed-radix (little-endian, \(L = [L_0,\dots,L_{D-1}]\)). Dequantization
is performed channel-wise into \([-1, 1]\) without residual
dependencies:

\[
\text{code}_d = \Bigl\lfloor \tfrac{k}{\prod_{j<d} L_j} \Bigr\rfloor \bmod L_d,
\qquad
x_d = \frac{\text{code}_d - \lfloor L_d/2 \rfloor}{\lfloor L_d/2 \rfloor}.
\]

For the entire codec, 8 codebooks with \(L = [8,7,6,6]\) yield
\(C = 32\) independent channels after unfolding, with \(L_k\) cyclically
matching \([8,7,6,6]\) (Equation A.1).

\hypertarget{a.2.-audio-interface-initialization}{%
\subsection{A.2. Audio Interface:
Initialization}\label{a.2.-audio-interface-initialization}}

During audio embedding initialization (§2.4), the following
transformations are performed:

\[
E_k \sim \mathcal{N}(0,1), \qquad
W_i \sim \mathcal{N}\!\bigl(0,\, \text{in}^{-1/2}\bigr), \quad b_i = 0, \qquad
s_\text{audio} \leftarrow \operatorname{std}(\text{embed\_tokens}).
\]

Before the heads, the prefix slice of length \(K\) is discarded,
aligning the hidden states of the backbone with the temporal dimension
of the audio region: \(h_t = \text{backbone}(x)_{K+t}\).

\hypertarget{a.3.-output-heads-and-target-functions}{%
\subsection{A.3. Output Heads and Target
Functions}\label{a.3.-output-heads-and-target-functions}}

The loss calculation for the 32 encoding heads (cross-entropy with a
causal shift) and the stop head (weighted binary cross-entropy) is
executed as follows (§2.5):

Let \(y_{t+1}^{(c)}\) be the target token for channel \(c \in \{1..C\}\)
(\(C=32\)) at step \(t+1\), and \(p_{t, v}^{(c)}\) be the predicted
probability of selecting token \(v\) at step \(t\). Let
\(\mathcal{T}_c = \{t \in [0, T-2] \mid y_{t+1}^{(c)} \neq \text{PAD}\}\)
be the set of valid time indices (excluding padding tokens
\(\text{PAD}\)):

\[\mathcal{L}_\text{CE} = \sum_{c=1}^{C} \frac{1}{|\mathcal{T}_c|}\sum_{t \in \mathcal{T}_c} - \log p_{t, y_{t+1}^{(c)}}^{(c)}.\]

For the stop head, let \(s_{t+1} \in \{0, 1\}\) be the true label
indicating the presence of a phantom stop frame at step \(t+1\), and
\(q_t\) be the output logit of the stop head at step \(t\). The binary
cross-entropy with a positive class weight \(w_{\text{pos}} = 25.0\) is:

\[\mathcal{L}_\text{stop} = \frac{1}{T-1}\sum_{t=0}^{T-2} - \left[ w_{\text{pos}} s_{t+1} \log \sigma(q_t) + (1 - s_{t+1}) \log(1 - \sigma(q_t)) \right],\]

where \(\sigma(\cdot)\) is the standard logistic sigmoid function. The
complete core loss function is:

\[\mathcal{L}_\text{core} = \mathcal{L}_\text{CE} + w_{\text{stop}}\mathcal{L}_\text{stop}, \qquad w_{\text{stop}} = 2.0.\]

To prevent division by zero (NaN-guard), if the set of valid indices is
empty for a channel \(c\) (\(|\mathcal{T}_c| = 0\)), the corresponding
cross-entropy term \(\mathcal{L}_{\text{CE}, c}\) is replaced with
\(0 \cdot \sum \text{logits}_c\), which zeroes the gradient while
preserving the autograd computation graph.

\hypertarget{a.4.-voice-cloning-compressor-cfg-and-regularizers}{%
\subsection{A.4. Voice Cloning: Compressor, CFG, and
Regularizers}\label{a.4.-voice-cloning-compressor-cfg-and-regularizers}}

The reference and prefix features (§2.6) are calculated as:

\[
\text{ref\_feats} = \mathrm{Linear}(C_\text{total} \to d)\bigl(\mathrm{dequant}(\mathrm{unfold}(\text{ref\_codes}))\bigr) + \mathrm{PE}_\text{sin},
\]

\[
\text{prefix} = \text{output\_scale}\cdot\mathrm{RMSNorm}(q), \qquad \text{output\_scale}_\text{init} = 1/\sqrt{d},
\]

where \(K = 8\) learnable queries pass through \(L = 2\) blocks
(self-attention → masked cross-attention → SwiGLU FFN, pre-norm
RMSNorm). CFG-dropout: with a probability of 0.15 (or forced for
null-sentinels), \(\text{prefix} \leftarrow \text{null\_prefix}\).

The compressor regularizers are calculated on the normalized
representation \(q_\text{normed}\) (\(\text{RMS} = 1\)) before the
curriculum masking stage. Diversity Loss (hinge-variance):

\[\mathcal{L}_\text{div} = \frac{1}{K}\sum_{j=1}^K \mathrm{relu}\bigl(\gamma - \operatorname{std}_K(q_\text{normed})\bigr), \qquad \gamma = 0.5.\]

SupCon on representation
\(z_i = \operatorname{norm}\bigl(\mathrm{MLP}(\operatorname{mean}_K(q_\text{normed}))\bigr)\)
with temperature \(\tau = 0.1\), related to VICReg
(\protect\hyperlink{ref-bardes2021vicreg}{Bardes, Ponce, and LeCun
2021}):

\[\mathcal{L}_\text{supcon} = \frac{1}{|A|}\sum_{i\in A} \frac{-1}{|P(i)|}\sum_{p\in P(i)} \log \frac{\exp(\langle z_i, z_p \rangle /\tau)}{\sum_{a\in V, a\neq i}\exp(\langle z_i, z_a \rangle /\tau)},\]

where \(V\) is the set of all valid (active, non-sentinel) samples in
the batch, \(P(i) \subseteq V \setminus \{i\}\) represents positive
examples (same speaker as \(i\)), and \(A \subseteq V\) is the set of
active anchors (\(|P(i)| \geq 1\)). The batch is formed via the
\(P\cdot K + M = 16\cdot 3 + 16\) scheme, and negative examples are
expanded using cross-rank merging.

Figure \ref{fig:refcompressor} summarizes the compressor data flow and
the two output taps: the \texttt{prefix} consumed by the decoder
(inference path) and the \texttt{q\_normed} representation feeding the
diversity and SupCon regularizers (training-only --- the compressor is
frozen during fine-tuning and DPO).

\hypertarget{a.5.-effective-rank-and-embedding-drift}{%
\subsection{A.5. Effective Rank and Embedding
Drift}\label{a.5.-effective-rank-and-embedding-drift}}

For weights \(W\) with singular values \(s\) (§3.1):

\[
p_i = \frac{s_i}{\sum_j s_j}, \qquad \operatorname{eff\_rank}(W) = \exp\!\Bigl(-\sum_i p_i \log p_i\Bigr).
\]

For the audio table \(E_k \in \mathbb{R}^{L_k \times 32}\), the number
of singular values equals \(L_k\), making the effective rank
structurally bounded: \(\operatorname{eff\_rank} \le L_k\). Text
embedding drift relative to initialization is:

\[
\text{drift} = \frac{\lVert W_\text{now} - W_\text{init}\rVert_F}{\lVert W_\text{init}\rVert_F}.
\]

\hypertarget{a.6.-dpo}{%
\subsection{A.6. DPO}\label{a.6.-dpo}}

Log-likelihood calculation of trajectory \(y\) of length \(T\) frames
given conditional input \(x\) (32 heads and Bernoulli stop-probability
components) with clipping
\(p_\text{stop} \leftarrow \max(p_\text{stop}, p_\text{floor})\),
\(p_\text{floor} = 10^{-4}\) (§5.4.2):

\[\log\pi(y \mid x) = \sum_{t=0}^{T-1} \sum_{c=1}^{C} \log p_c\bigl(y_t^{(c)}\bigr) + \sum_{t=0}^{T-2}\log(1 - p_{\text{stop},t}) + \mathbb{I}(\text{not truncated}) \log p_{\text{stop},T-1}.\]

Length-normalized Bradley--Terry objective (\(\beta = 2.0\))
(\protect\hyperlink{ref-meng2024simpo}{Meng, Xia, and Chen 2024}):

\[\mathcal{L}_\text{DPO} = -\log\sigma\!\left(\beta\!\left[\frac{\log\pi_\theta(y_w \mid x) - \log\pi_\text{ref}(y_w \mid x)}{T_w} - \frac{\log\pi_\theta(y_l \mid x) - \log\pi_\text{ref}(y_l \mid x)}{T_l}\right]\right),\]

where \(y_w\) and \(y_l\) are the selected (winning) and rejected
(losing) trajectories with lengths \(T_w\) and \(T_l\), respectively.

Reward function \(R(y)\) for automatic pair ranking (two-sided length
and quality penalty):

\[
\begin{aligned}
R(y) = &- w_\text{wer} \operatorname{WER}(y) - w_\text{over} \max\bigl(0, d(y) - d_\text{max}\bigr) - w_\text{short} \mathbb{I}\bigl(d(y) < d_\text{min}\bigr) \\
       &- w_\text{empty} \mathbb{I}(\text{ASR is empty}) - w_\text{sim} \bigl(1 - \cos(y)\bigr),
\end{aligned}
\]

where \(d(y)\) is the synthesized speech duration in seconds,
\(d_{\text{min}}\) and \(d_{\text{max}}\) are acceptable duration
boundaries, \(\operatorname{WER}(y)\) is the Whisper word error rate,
and \(\cos(y)\) is the cosine similarity of the generated and target
speaker embeddings.

\hypertarget{a.7.-speed-metrics}{%
\subsection{A.7. Speed Metrics}\label{a.7.-speed-metrics}}

Speed metrics are calculated as follows (§6.2):

\[
\text{RTF} = \frac{\text{wall}_i}{\text{audio}_i}\ (\text{per-stream};\ <1 \Rightarrow \text{faster than real-time}), \qquad
\text{xRT} = \frac{\sum_i \text{audio}_i}{\text{wall}_{\text{elapsed}}}.
\]

Here, \(\text{wall}_i\) and \(\text{audio}_i\) represent the individual
request wall duration and generated audio length for stream \(i\),
respectively. \(\text{wall}_{\text{elapsed}}\) is the total elapsed time
of the concurrent batch run. Streams with
\(\text{audio}_i < 2.0\text{ s}\) are filtered out to prevent stochastic
early stops from biasing the percentiles.

All concurrency measurements are conducted over local loopback
(\texttt{localhost}), thus excluding external network transmission
overhead. The request traffic model simulates an instantaneous burst
(load spike) where all \(C\) streams are initiated concurrently,
representing a worst-case contention scenario for the prefill phase,
rather than a distributed Poisson arrival process.

\(\text{TTFB}\) represents the wall time elapsed to the first audio
delta (equivalent to TTFA).

\begin{center}\rule{0.5\linewidth}{0.5pt}\end{center}

\hypertarget{appendix-b.-development-notes-scale-mismatches-and-gradient-attenuation-diagnostics}{%
\section{Appendix B. Development Notes: Scale Mismatches and Gradient
Attenuation
Diagnostics}\label{appendix-b.-development-notes-scale-mismatches-and-gradient-attenuation-diagnostics}}

The architectural and optimization choices presented in the main text
were derived from targeted empirical diagnostics during the development
phase. This appendix documents key numerical observations, scale
mismatches, and gradient issues encountered during early training
iterations.

\hypertarget{b.1.-modality-scale-mismatch}{%
\subsection{B.1. Modality Scale
Mismatch}\label{b.1.-modality-scale-mismatch}}

In early pretraining runs, a severe representation scale mismatch was
observed between the text and audio modalities at the input of the
transformer backbone: * \textbf{Pretrained Text Embeddings:} The text
embedding vectors of the pretrained Qwen3.5 base model had an average
\(L_2\) norm of \(\|x_{\text{text}}\|_2 \approx 0.5\text{--}0.7\)
(corresponding to a root-mean-square magnitude of
\(\text{RMS} \approx 0.016\) for hidden dimension \(d = 1024\)). *
\textbf{Standard Audio Embeddings:} Randomly initialized embedding
lookup tables \(E_c\) under standard initialization
\(\mathcal{N}(0, 1)\) yielded an expected \(L_2\) norm of
\(\sqrt{d} = \sqrt{1024} \approx 32\).

Direct concatenation of these unaligned modalities led to a \(60\times\)
scale mismatch. In the self-attention mechanism, the dot products of
query and key vectors were dominated by the audio tokens:
\[\text{score}(\text{audio}_q, \text{audio}_k) \propto 32 \times 32 = 1024\]
\[\text{score}(\text{audio}_q, \text{text}_k) \propto 32 \times 0.7 \approx 22.4\]

Consequently, after the softmax operation, the text tokens received
\(\approx 50\times\) less attention weight. The model structurally
ignored the text context and optimized the objective solely as an
audio-only autoregressive generator, rendering the text conditioning
ineffective.

\hypertarget{b.2.-gradient-attenuation-path}{%
\subsection{B.2. Gradient Attenuation
Path}\label{b.2.-gradient-attenuation-path}}

To align the forward-pass scales, early configurations utilized a
scaling multiplier \(s_{\text{audio}} = 0.02\) following a shared audio
projection LayerNorm. However, tracking the backpropagation path from
the loss \(\mathcal{L}\) to the audio embedding weights
\(W_{\text{emb}}\) revealed a severe gradient starvation effect. The
chain rule for the gradient is:

\[\frac{\partial \mathcal{L}}{\partial W_{\text{emb}}} = \frac{\partial \mathcal{L}}{\partial h_{\text{backbone}}} \cdot \frac{\partial h_{\text{backbone}}}{\partial x_{\text{audio}}} \cdot s_{\text{audio}} \cdot \mathbf{J}_{\text{RMSNorm}} \cdot W_{\text{proj}}^T \cdot \mathbf{X}_{\text{lookup}}\]

where: 1. \textbf{Forward Scale:} The scaling multiplier
\(s_{\text{audio}} = 0.02\) caused a \(50\times\) attenuation of the
gradient in the backward pass. 2. \textbf{RMSNorm Jacobian:} The
Jacobian of the RMSNorm operation divides the incoming gradient by the
RMS of the projected sum (which grew to \(\approx 1.47\)), attenuating
it by \(1.47\times\). Additionally, the projection operator
\((\mathbf{I} - \hat{x}\hat{x}^T)\) forced the gradient updates of
different tokens to align in direction, causing gradient homogenization.
3. \textbf{Lookup Factor:} The embedding lookup backpropagation
introduced an additional \(1/\sqrt{d} \approx 1/32 \approx 0.031\times\)
scaling.

The combined gradient attenuation factor \(\gamma\) was:
\[\gamma \approx 0.02 \times \frac{1}{1.47} \times \frac{1}{32} \approx 0.00042\]

The gradient norm reaching the audio embedding tables was
\(\approx 2400\) times weaker than the gradient norm of the backbone
hidden states. The embedding tables remained practically frozen, and all
representation learning was forced into the projection layers. This
bottleneck was eliminated in later pretraining iterations by removing
the intermediate RMSNorm, moving the scaling factor, and initializing
the audio tables directly with the pretrained text embedding variance
\(\mathcal{N}(0, \text{text\_std})\).

\hypertarget{b.3.-representational-alignment-in-hidden-space}{%
\subsection{B.3. Representational Alignment in Hidden
Space}\label{b.3.-representational-alignment-in-hidden-space}}

In early runs, weight-space diagnostics showed that the cosine
similarity between the text embedding matrix and the audio embedding
matrix remained close to zero, suggesting that the two modalities
remained orthogonal at the input.

However, monitoring activation states via forward hooks (adjusted for
FSDP compatibility) revealed that the backbone transformer successfully
aligned the modalities in its deeper layers. While the cosine similarity
between text and audio states at Layer 0 (input) was \(-0.04\), it rose
to \(+0.37\) at Layer 4, \(+0.59\) at Layer 7, and stabilized at
\textbf{\(0.98\text{--}0.99\)} at Layers 12 and 13. This confirmed that
the full-attention backbone acts as a powerful cross-modal aligner in
its deep latent space, even when the input embeddings are orthogonal.

\hypertarget{b.4.-generation-variance-in-short-phrases}{%
\subsection{B.4. Generation Variance in Short
Phrases}\label{b.4.-generation-variance-in-short-phrases}}

To determine whether the high Word Error Rate (WER) on short phrases was
due to a capacity limitation (the model not knowing how to synthesize
specific short words) or generation instability (high variance), we
analyzed the distribution of rewards across multiple rollout groups.

On the on-policy diagnostic set of the second DPO round (20 rollouts per
prompt-voice pair), we evaluated the minimum WER achieved within each
group. In 100\% of the failure-prone groups (where the mean success rate
was below 50\%), the minimum WER within the group was \textbf{\(0.00\)}.
This proved that the model had the physical capability to synthesize
every word in the stress test but suffered from high generation
variance, validating the preference optimization (DPO) framework as the
correct lever: contract the policy distribution around the successful
mode rather than collect more training data.

\begin{center}\rule{0.5\linewidth}{0.5pt}\end{center}

\hypertarget{appendix-c.-comparison-against-commercial-and-large-scale-systems}{%
\section{Appendix C. Comparison Against Commercial and Large-Scale
Systems}\label{appendix-c.-comparison-against-commercial-and-large-scale-systems}}

Beyond the open-source voice-cloning cohort (§6.3), we ran five
large-scale and commercial systems on the same 1088 Seed-TTS-eval
prompts: Cartesia, ElevenLabs, Grok-TTS, Kokoro, and VibeVoice. These
were evaluated \textbf{without reference cloning}
(\texttt{seed\_tts\_eval\_en} mode), so SIM is undefined for them and
the comparison is restricted to intelligibility (WER/CER) and
signal-quality metrics (UTMOS, NISQA-MOS, NOI, COL, DIS), which use the
identical evaluation stack and remain directly comparable. (Kokoro and
VibeVoice are themselves open-weights; we group them here with the
commercial systems because, like them, they were run in the non-cloning
\texttt{en} mode on this evaluation.)

We separate this comparison from the main results deliberately: (i) it
is not voice cloning --- no SIM, a different task setup; and (ii) Gepard
does not target SOTA intelligibility (§1.4). The purpose is to situate
the full field, not to claim parity. Gepard is shown in the table below
for reference.

\begin{table}[h]
\centering
\footnotesize
\setlength{\tabcolsep}{6pt}
\renewcommand{\arraystretch}{1.15}
\begin{tabular}{lcccccc}
\hline
\textbf{Model} & WER\,$\downarrow$ & UTMOS\,$\uparrow$ & NISQA-MOS\,$\uparrow$ & NOI\,$\uparrow$ & COL\,$\uparrow$ & DIS\,$\uparrow$ \\
\hline
ElevenLabs & \textbf{0.014} & 2.95 & 4.78 & 4.27 & 4.40 & 4.61 \\
Grok-TTS & 0.015 & 3.20 & \textbf{5.10} & \textbf{4.65} & \textbf{4.63} & \textbf{4.85} \\
Cartesia & 0.015 & \textbf{3.39} & 4.82 & 4.47 & 4.56 & 4.75 \\
Kokoro & 0.017 & 3.18 & 5.05 & 4.64 & 4.59 & 4.77 \\
\hdashline
\textbf{Gepard (Ours)} & 0.036 & 2.64 & 4.25 & 4.16 & 4.16 & 4.51 \\
\hdashline
VibeVoice & 0.109 & 2.93 & 4.17 & 3.69 & 4.08 & 4.30 \\
\hline
\end{tabular}
\end{table}

Sorted by WER; best value in each column is bold. SIM is omitted (not
defined in the non-cloning mode).

The commercial systems form a tight, high-quality cluster: WER
\(\approx 0.014\text{--}0.017\) with zero hard failures, and the highest
NISQA-MOS in the entire field (Grok-TTS 5.10, Kokoro 5.05, Cartesia
4.82, ElevenLabs 4.78). Gepard's WER (0.036) sits below this cluster but
well above VibeVoice (0.109), and its NISQA-MOS (4.25), while behind the
commercial leaders, remains clean and stable. We do not claim parity
with these systems; the comparison establishes the upper envelope of the
field and confirms Gepard's positioning as a fast, stable system rather
than a SOTA-intelligibility one.

\begin{figure}
\centering
\includegraphics[width=1\textwidth,height=\textheight]{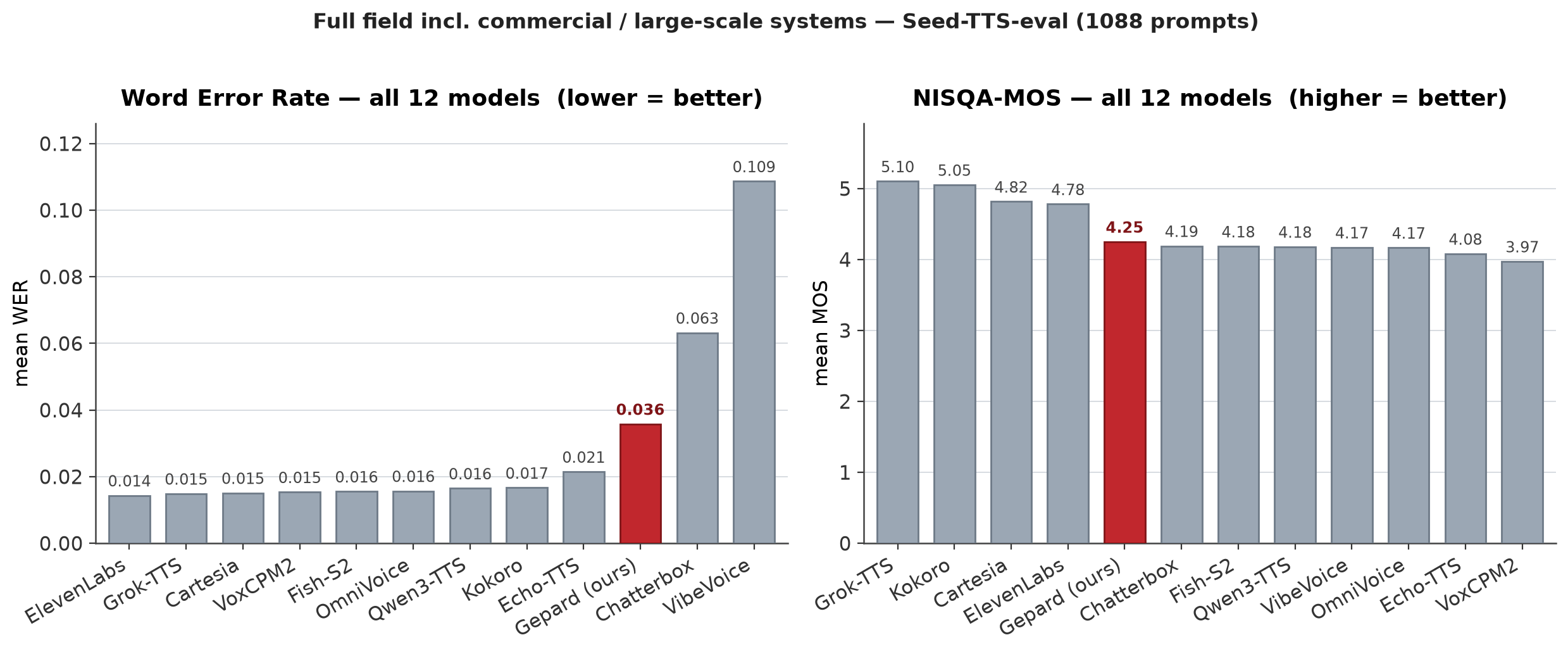}
\caption{Full field including commercial / large-scale systems on
Seed-TTS-eval (1088 prompts): WER\(\downarrow\) (left) and
NISQA-MOS\(\uparrow\) (right) across all 12 models. The commercial
cluster leads on both axes; Gepard (crimson) is mid-field on WER and the
best of the open-source voice-cloning cohort on
NISQA-MOS.\label{fig:bench-appendix-field}}
\end{figure}

\begin{center}\rule{0.5\linewidth}{0.5pt}\end{center}

\hypertarget{references}{%
\section*{References}\label{references}}
\addcontentsline{toc}{section}{References}

\hypertarget{refs}{}
\begin{CSLReferences}{1}{0}
\leavevmode\vadjust pre{\hypertarget{ref-alayrac2022flamingo}{}}%
Alayrac, Jean-Baptiste, Jeff Donahue, Pauline Luc, Antoine Miech,
Jean-Baptiste Barr, Yana Hasson, Diane Larlus, Andrew Zisserman, et al.
2022. {``Flamingo: A Visual Language Model for Few-Shot Learning.''}
\emph{arXiv Preprint arXiv:2204.14198}.

\leavevmode\vadjust pre{\hypertarget{ref-bytedance2024seedtts}{}}%
Anastassiou, Philip et al. 2024. {``Seed-TTS: A Family of High-Quality
Zero-Shot Text-to-Speech Models.''} \emph{arXiv Preprint
arXiv:2406.02430}.

\leavevmode\vadjust pre{\hypertarget{ref-bardes2021vicreg}{}}%
Bardes, Adrien, Jean Ponce, and Yann LeCun. 2021. {``VICReg:
Variance-Invariance-Covariance Regularization for Self-Supervised
Learning.''} \emph{arXiv Preprint arXiv:2105.04906}.

\leavevmode\vadjust pre{\hypertarget{ref-nvidia2025nanocodec}{}}%
Casanova, Edresson, Paarth Neekhara, Ryan Langman, Shehzeen Hussain,
Subhankar Ghosh, Xuesong Yang, Ante Jukić, Jason Li, and Boris Ginsburg.
2025. {``NanoCodec: Towards High-Quality Ultra Fast Speech LLM
Inference.''} \emph{arXiv Preprint arXiv:2508.05835}.

\leavevmode\vadjust pre{\hypertarget{ref-chen2021wavlm}{}}%
Chen, Sanyuan, Chengyi Wang, Zhengyang Chen, Yu Wu, Shujie Liu, Zhuoyuan
Chen, Jinyu Li, et al. 2021. {``WavLM: Large-Scale Self-Supervised
Pre-Training for Full Stack Speech Processing.''} \emph{arXiv Preprint
arXiv:2110.13900}.

\leavevmode\vadjust pre{\hypertarget{ref-copet2023musicgen}{}}%
Copet, Jade, Felix Kreuk, Itai Gat, Glen Talgorn, Gabriel Synnaeve,
Yossi Adi, and Alexandre Défossez. 2023. {``Simple and Controllable
Music Generation.''} \emph{arXiv Preprint arXiv:2306.05284}.

\leavevmode\vadjust pre{\hypertarget{ref-dao2023flashattention2}{}}%
Dao, Tri. 2023. {``FlashAttention-2: Faster Attention with Better
Parallelism and Work Partitioning.''} \emph{arXiv Preprint
arXiv:2307.08691}.

\leavevmode\vadjust pre{\hypertarget{ref-defossez2024moshi}{}}%
Défossez, Alexandre et al. 2024. {``Moshi: A Speech-Text Foundation
Model for Real-Time Dialogue.''} \emph{arXiv Preprint arXiv:2410.00037}.

\leavevmode\vadjust pre{\hypertarget{ref-defossez2022encodec}{}}%
Défossez, Alexandre, Jade Copet, Gabriel Synnaeve, and Yossi Adi. 2022.
{``High Fidelity Neural Audio Compression.''} \emph{arXiv Preprint
arXiv:2210.13438}.

\leavevmode\vadjust pre{\hypertarget{ref-du2024vallt}{}}%
Du, Chenpeng, Yiwei Guo, Feiyu Shen, Kai Yu, and Daniel P. Povey. 2024.
{``VALL-t: Decoder-Only Generative Transducer for Robust and
Decoding-Controllable Text-to-Speech.''} \emph{arXiv Preprint
arXiv:2401.14321}.

\leavevmode\vadjust pre{\hypertarget{ref-framestacked2025}{}}%
Fejgin, Roy, Paarth Neekhara, Shehzeen Hussain, Subhankar Ghosh, Mikyas
T. Desta, Rafael Valle, Jason Li, and Boris Ginsburg. 2025.
{``Frame-Stacked Local Transformers for Efficient Multi-Codebook Speech
Generation.''} \emph{arXiv Preprint arXiv:2509.19592}.

\leavevmode\vadjust pre{\hypertarget{ref-ho2022cfg}{}}%
Ho, Jonathan, and Tim Salimans. 2022. {``Classifier-Free Diffusion
Guidance.''} \emph{arXiv Preprint arXiv:2207.12598}.

\leavevmode\vadjust pre{\hypertarget{ref-koeltts2025}{}}%
Hussain, Shehzeen, Paarth Neekhara, Xuesong Yang, Edresson Casanova,
Subhankar Ghosh, Mikyas T. Desta, Roy Fejgin, Rafael Valle, and Jason
Li. 2025. {``Koel-TTS: Enhancing LLM Based Speech Generation with
Preference Alignment and Classifier Free Guidance.''} \emph{arXiv
Preprint arXiv:2502.05236}.

\leavevmode\vadjust pre{\hypertarget{ref-jaegle2021perceiverio}{}}%
Jaegle, Andrew, Sebastian Borgeaud, Jean-Baptiste Alayrac, Carl Doersch,
Catalin Ionescu, David Ding, Skanda Koppula, et al. 2021. {``Perceiver
IO: A General Architecture for Structured Inputs \& Outputs.''} In
\emph{arXiv Preprint arXiv:2107.14795}.

\leavevmode\vadjust pre{\hypertarget{ref-khosla2020supcon}{}}%
Khosla, Prannay, Piotr Teterwak, Chen Wang, Aaron Sarna, Yonglong Tian,
Phillip Isola, Aaron Maschinot, Ce Liu, and Dilip Krishnan. 2020.
{``Supervised Contrastive Learning.''} In \emph{Advances in Neural
Information Processing Systems (NeurIPS)}, 33:18661--73.
\url{https://arxiv.org/abs/2004.11362}.

\leavevmode\vadjust pre{\hypertarget{ref-kumar2022lpft}{}}%
Kumar, Ananya, Aditi Raghunathan, Robbie Jones, Tengyu Ma, and Percy
Liang. 2022. {``Fine-Tuning Can Distort Pretrained Features and
Underperform Out-of-Distribution.''} \emph{arXiv Preprint
arXiv:2202.10054}.

\leavevmode\vadjust pre{\hypertarget{ref-kumar2023dac}{}}%
Kumar, Rithesh et al. 2023. {``High-Fidelity Audio Compression with
Improved Descript Audio Codec.''} \emph{arXiv Preprint
arXiv:2306.06546}.

\leavevmode\vadjust pre{\hypertarget{ref-kwon2023vllm}{}}%
Kwon, Woosuk, Zhuohan Li, Siyuan Zhuang, Ying Sheng, Lianmin Zheng, Cody
Hao Yu, Joseph E Gonzalez, Hao Zhang, and Ion Stoica. 2023. {``Efficient
Memory Management for Large Language Model Serving with
PagedAttention.''} In \emph{Proceedings of the 29th Symposium on
Operating Systems Principles (SOSP)}, 611--26.
\url{https://arxiv.org/abs/2309.06180}.

\leavevmode\vadjust pre{\hypertarget{ref-li2023blip2}{}}%
Li, Junnan, Dongxu Li, Silvio Savarese, and Steven Hoi. 2023. {``BLIP-2:
Bootstrapping Language-Image Pre-Training with Frozen Image Encoders and
Large Language Models.''} \emph{arXiv Preprint arXiv:2301.12597}.

\leavevmode\vadjust pre{\hypertarget{ref-meng2024simpo}{}}%
Meng, Yu, Mengzhou Xia, and Danqi Chen. 2024. {``SimPO: Simple
Preference Optimization with a Reference-Free Reward.''} \emph{arXiv
Preprint arXiv:2405.14734}.

\leavevmode\vadjust pre{\hypertarget{ref-mentzer2023fsq}{}}%
Mentzer, Fabian, Eirikur Agustsson, Michael Tschannen, Srikanth
Malireddy, and Elena Alshina. 2023. {``Finite Scalar Quantization:
VQ-VAE Made Simple.''} \emph{arXiv Preprint arXiv:2309.15505}.

\leavevmode\vadjust pre{\hypertarget{ref-mittag2021nisqa}{}}%
Mittag, Gabriel et al. 2021. {``NISQA: A Deep Learning-Based Model for
Speech Quality Assessment.''} \emph{arXiv Preprint arXiv:2104.09494}.

\leavevmode\vadjust pre{\hypertarget{ref-nvidia2024magpie}{}}%
Neekhara, Paarth, Shehzeen Hussain, Subhankar Ghosh, Jason Li, Rafael
Valle, Rohan Badlani, and Boris Ginsburg. 2024. {``Improving Robustness
of LLM-Based Speech Synthesis by Learning Monotonic Alignment.''}
\emph{arXiv Preprint arXiv:2406.17957}.

\leavevmode\vadjust pre{\hypertarget{ref-voicestar2025}{}}%
Peng, Puyuan, Shang-Wen Li, Abdelrahman Mohamed, and David Harwath.
2025. {``VoiceStar: Robust Zero-Shot Autoregressive TTS with Duration
Control and Extrapolation.''} \emph{arXiv Preprint arXiv:2505.19462}.

\leavevmode\vadjust pre{\hypertarget{ref-radford2022whisper}{}}%
Radford, Alec, Jong Wook Kim, Tao Xu, Greg Brockman, Christine McLeavey,
and Ilya Sutskever. 2022. {``Robust Speech Recognition via Large-Scale
Weak Supervision.''} \emph{arXiv Preprint arXiv:2212.04356}.

\leavevmode\vadjust pre{\hypertarget{ref-rafailov2023dpo}{}}%
Rafailov, Rafael, Archit Sharma, Eric Mitchell, Stefano Ermon,
Christopher D Manning, and Chelsea Finn. 2023. {``Direct Preference
Optimization: Your Language Model Is Secretly a Reward Model.''}
\emph{arXiv Preprint arXiv:2305.18290}.

\leavevmode\vadjust pre{\hypertarget{ref-roy2007effective}{}}%
Roy, Olivier, and Martin Vetterli. 2007. {``The Effective Rank: A
Measure of Effective Dimensionality.''} In \emph{2007 15th European
Signal Processing Conference (EUSIPCO)}, 606--10. IEEE.

\leavevmode\vadjust pre{\hypertarget{ref-saeki2022utmos}{}}%
Saeki, Takaaki et al. 2022. {``UTMOS: UTokyo-SaruLab System for VoiceMOS
Challenge 2022.''} In \emph{Interspeech}.
\url{https://arxiv.org/abs/2204.02152}.

\leavevmode\vadjust pre{\hypertarget{ref-shao2024deepseekmath}{}}%
Shao, Zhihong, Peiyi Wang, Qihao Zhu, Runxin Xu, Junxiao Song, Mingchuan
Zhang, Y. K. Li, Y. Wu, and Daya Guo. 2024. {``{DeepSeekMath: Pushing
the Limits of Mathematical Reasoning in Open Language Models}.''}
\emph{arXiv Preprint arXiv:2402.03300}.

\leavevmode\vadjust pre{\hypertarget{ref-wang2023valle}{}}%
Wang, Chengyi, Sanyuan Zhou, Shujie Liu, Yu Chen, Yuxuan Wu, Shujie Liu,
Jinyu Li, et al. 2023. {``Neural Codec Language Models Are Zero-Shot
Text to Speech Synthesizers.''} \emph{arXiv Preprint arXiv:2301.02111}.

\leavevmode\vadjust pre{\hypertarget{ref-attentionguidance2025}{}}%
Wang, ShiMing, ZhiHao Du, Yang Xiang, TianYu Zhao, Han Zhao, Qian Chen,
XianGang Li, HanJie Guo, and ZhenHua Ling. 2025. {``Eliminating
Stability Hallucinations in LLM-Based TTS Models via Attention
Guidance.''} \emph{arXiv Preprint arXiv:2509.19852}.

\leavevmode\vadjust pre{\hypertarget{ref-wang2019multimodal}{}}%
Wang, Zihang et al. 2019. {``What Makes Multimodal Networks Hard to
Train?''} \emph{arXiv Preprint arXiv:1905.12681}.

\leavevmode\vadjust pre{\hypertarget{ref-xia2024valle2}{}}%
Xia, Ziyang, Sanyuan Zhou, Chengyi Wang, Yu Chen, Yuxuan Wu, Shujie Liu,
Jinyu Li, et al. 2024. {``VALL-e 2: Neural Codec Language Models Are
Human-Like Zero-Shot Text-to-Speech Synthesizers.''} \emph{arXiv
Preprint arXiv:2406.05370}.

\leavevmode\vadjust pre{\hypertarget{ref-yang2022mup}{}}%
Yang, Greg, Edward J Hu, Igor Babuschkin, Szymon Sidor, Xiaodong Liu,
David Farheit, et al. 2022. {``Tensor Programs v: Tuning Large Neural
Networks via Zero-Shot Hyperparameter Transfer.''} \emph{arXiv Preprint
arXiv:2203.03466}.

\end{CSLReferences}

\end{document}